\documentclass[sn-aps]{sn-jnl}

\jyear{2022}%

\theoremstyle{thmstyleone}%
\usepackage{grffile} 

\def\bfr{\mathbf{r}}
\def\bfv{\mathbf{v}}
\def\bfx{\mathbf{x}}

\def\bfk{\mathbf{k}}

\def\fs8{f\sigma_8}

\def\hinvmpc{\,h^{-1}{\rm Mpc}}
\def\Mpc{\,{\rm Mpc}}

\def\hmpcinv{\,h\,{\rm Mpc^{-1}}}
\def\Msun{M_\odot}

\def\kms{\,{\rm km\,s^{-1}}}

\newcommand{\omgeom}{\Omega_{\rm{M}}^{\rm geom}}
\newcommand{\omgrow}{\Omega_{\rm{M}}^{\rm grow}}
\newcommand{\om}{\Omega_{\rm M}}

\newcommand{\wgeom}{w^{\rm geom}}
\newcommand{\wgrow}{w^{\rm grow}}
\theoremstyle{thmstyletwo}%

\theoremstyle{thmstylethree}%

\begin{document}

\title[Growth of Structure]{Growth of Cosmic  Structure}


\author{\fnm{Dragan} \sur{Huterer}}

\affil{\orgdiv{Department of Physics and
    Leinweber Center for Theoretical Physics}, \orgname{University of Michigan},
  \orgaddress{\street{450 Church St}, \city{Ann Arbor, MI}, \postcode{48103},
    \country{USA}}}


\abstract{We review one of the most fruitful areas in cosmology today that
  bridge theory and data --- the temporal growth of large-scale structure. We
  go over the growth's physical foundations, and derive its behavior in
  simple cosmological models. While doing so, we explain how measurements of
  growth can be used to understand theory. We then review how some of the most
  mature cosmological probes --- galaxy clustering, gravitational lensing, the
  abundance of clusters of galaxies, cosmic velocities, and cosmic microwave
  background --- can be used to probe the growth of structure. We report
  the current constraints on growth, which are summarized as measurements of the
  parameter combination $\fs8$ as a function of redshift, or else as the mass
  fluctuation amplitude parameter $S_8$.  We finally illustrate several
  statistical approaches, ranging from the ``growth index'' parameterization
  to more general comparisons of growth and geometry, that can sharply test
  the standard cosmological model and  indicate the presence of
  modifications to general relativity.}




\maketitle

\section{Introduction}\label{sec1}

In the standard cosmological model, the seeds of structure laid out by
inflation were amplified by the influence of gravity. The 
density perturbation is defined as
\begin{equation}
  \delta(\bfx, t)\equiv \frac{\delta\rho(\bfx, t)}{\bar\rho}
\end{equation}
where $\rho$ is density in matter, and $\bfx$ and $t$ denote space and time
coordinates respectively. Density perturbations are seeded by inflation, 
start out with an amplitude $\delta\simeq 10^{-5}$, and are subsequently
amplified by gravity over the ensuing billions of years in a way that is
described by Eq.~(\ref{eq:growth_eq_with_cs_Fourier}) below. The general fact
that structure grows over time is a very well-established feature of the hot
big-bang cosmological model. The growth of structure can be measured through
observations of positions and motions of galaxies as a function of cosmic
time, and by comparing the inferred overdensities to the initial conditions,
$\delta\sim 10^{-5}$, that can be observed in the temperature anisotropies of
the cosmic microwave background (CMB).

A cosmological model predicts the temporal growth of structure. Such a
prediction is a key ingredient in theoretical calculations of the various
observable quantities that can be experimentally measured. For example, weak
lensing shear integrates the power spectrum of structure along the line of
sight (as we discuss in Sec.~\ref{sec:shear}), which is in turn is determined by
the growth of structure. As another example, the amplitude of galaxies'
peculiar velocities, as well as the spatial correlations of galaxy positions,
are both largely determined by the growth of structure. In all of these
examples, a larger rate of growth corresponds to a larger signal (assuming
fixed initial conditions). Turning the argument around, measurements of the
large-scale structure and inference of its statistical properties inform us
about the growth of structure. In turn, constraints on the growth of structure
help pin down the parameters of the cosmological model, and are especially
sensitive to the properties of dark energy.
 
Over the past two decades, the advent of massive new datasets from large-scale
structure has brought into focus the importance of the growth of
structure. This is principally because growth is a sensitive probe of both
dark energy and modified gravity, as hence its measurements can distinguish
between the two. Specifically, in the context of general relativity, growth is
mathematically related to the geometrical measures such as
  distances. Specifically, growth of structure affects the expansion rate
  given by the Hubble parameter $H$, which in turn affects distances (as $H$
  enters the distance-redshift relation). Therefore, growth is directly
  related to the geometrical quantities in the universe.  However, this
relation between growth and geometry assumes Einstein's general relativity, and is
generally broken (or modified) when gravity itself is modified. Comparing growth to geometry thus
enables stringent tests of modified gravity.

The purpose of this review is to present a rather high-level review of the
growth of structure, aimed at a non-expert astrophysicist.  Our main goal to
lay out the theoretical foundations for basic studies of growth. We also wish
to illustrate, at a basic yet quantitative level, how growth enters the various cosmological
probes, and how they can be used to measure growth. Finally,
we review some of the most promising strategies to use growth to better
understand dark energy and modified gravity. This review focuses on the big-picture
foundations of the subject and the cosmological utility of the growth of
structure (as well as current constraints), and updates and  complements  the previous
major reviews of the subject \cite{Weinberg:2013agg,Huterer:2013xky} which focused on future
surveys and expected systematic errors in the associated cosmological probes.

This review is organized as follows. In Sec.~\ref{sec:theory} we discuss the
theoretical background useful for understanding growth of structure. in
Sec.~\ref{sec:obs}, we review the observations that are sensitive to
growth, and compile recent constraints on the growth of
structure. In Sec.~\ref{sec:consistency}, we discuss ways in which the growth
of structure can be used to probe dark energy and general relativity. We conclude in Sec.~\ref{sec:concl}.

\section{Theoretical background}\label{sec:theory}

Here we outline the theoretical predictions for the growth of cosmic
structure. We start with  density perturbations, introduce the (linear) growth
of structure, derive an equation that governs its evolution in time, and study
its solutions.

\subsection{Temporal evolution of linear density perturbations}

In our theoretical treatment, we will specialize in linear
theory, and assume small fluctuations, with $\lvert\delta\rvert \ll 1$.
To make even better progress, we expand the overdensity in Fourier
basis.  We do so because it turns out that each Fourier mode $\delta_\bfk$
evolves independently (assuming linear perturbations and standard general
relativity).
Following a standard,
non-relativistic, perturbation-theory analysis that combines the continuity,
Euler, and Poisson equations, one arrives at (e.g.\ \citep{2010gfe..book.....M,Huterer:2023mmv})
\begin{equation}
\frac{\partial^2\delta_{\bf k}}{\partial t^2} + 
2\frac{\dot{a}}{a} \frac{\partial\delta_{\bf k}}{\partial t} =
\left (4\pi G \rho_{\rm M} - \frac{k^2 c_s^2}{a^2}\right )\delta_{\rm k}
- \frac{2}{3}\frac{T}{a^2}k^2 S_{\bf k},
\label{eq:growth_eq_with_cs_Fourier}
\end{equation}
where $T$ is temperature, $a$ is the scale factor, $c_s$ is the speed of
sound, and $\rho_{\rm M}$ is the mean matter density; all of these quantities depend
on cosmic time $t$.  Further, $\delta_{\bf k}$ is the mode in the Fourier
expansion of the overdensity field at some wavenumber $\bfk$
\begin{equation}
\delta_{\bf k}(t) = \frac{1}{\sqrt{V}}\int
\delta({\bf r},t)e^{-i{\bf k}\cdot{\bf r}}d^3{\bf r}.
\end{equation}
Here $V$ is the volume of the larger region over which the perturbations are
assumed to be periodic (note that ${\bf k}$ and ${\bf r}$ are both comoving
quantities).  Similarly, $S_\bfk$ is the Fourier mode of entropy perturbations.

We further specialize in isentropic initial conditions where there is no
fluctuation in entropy in the initial conditions, so that $S_\bfk\propto
\nabla S = 0$. [Confusingly, these fluctuations are most often called
  adiabatic initial conditions, which strictly speaking implies $\dot{S}=0$,
  not $\nabla S=0$; we henceforth adopt this imprecise but popular
  nomenclature.] In the presence of adiabatic initial conditions, fluctuations
in various components (matter, radiation, neutrinos etc) are proportional to
each other, and the overall curvature fluctuation is nonzero. This is the kind
of initial condition that inflation typically predicts, and that current data
favor. Finally, we specialize in fluctuations on scales larger than the Jeans
scale, so $k\ll k_{\rm Jeans}\equiv \sqrt{4\pi G \rho_{\rm M}}(a/c_s)$. This is a
reasonable assumption since, after recombination, the Jeans scale corresponds
to mass of order $10^5\Msun$, which is much smaller than the structures that
we will be interested in which are roughly mass of a galaxy or larger, so
$\gtrsim 10^{12}\Msun$. In this regime, we can drop the term that is
proportional to $k^2$. With all that, Eq.~(\ref{eq:growth_eq_with_cs_Fourier})
simplifies to
\begin{equation}
\ddot\delta + 2H\dot\delta-4\pi G\rho_{\rm M} \delta = 0,
\label{eq:growth}
\end{equation}
where $H\equiv \dot{a}/a$ is the Hubble parameter.
Notice that we have dropped the subscript $_\bfk$ in the overdensity, as there is no more
wavenumber dependence in this equation.

Equation (\ref{eq:growth}) describes the evolution of the density
perturbations on scales $ 0.001\hmpcinv \lesssim k\lesssim 0.1\hmpcinv$, which
is where the most usable observational data reside. [On scales larger than
  about $0.001\hmpcinv$ there are additional general-relativistic corrections
  and may anyway difficult to probe, while scales smaller than about $0.1\hmpcinv$ lie in
  the nonlinear regime;  see the bullet points below for further discussion.]
This is a second-order ordinary differential equation
for the matter overdensity $\delta$. We will solve it below to get solutions
for the linear growth of structure in simple cosmological models.

Let us first review  the assumptions that were assumed in deriving
 Eq.~(\ref{eq:growth}):
\begin{itemize}
\item General theory of relativity;\\[-0.3cm]
\item Adiabatic initial conditions;\\[-0.3cm]
\item Sub-horizon scales (i.e.\ relativistic effects are ignored); $k\gg H_0$;\\[-0.3cm]
\item Spatial scales above the Jeans length ($k\ll k_J$);\\[-0.3cm]
\item Linear theory ($\delta\ll 1$), which corresponds roughly to scales
  $k\lesssim 0.1\hmpcinv$ today.
\end{itemize}
The breakdown of any of these assumptions could lead to the invalidation of
Eq.~(\ref{eq:growth}), which would be manifested as a difference between the
predicted and observed growth. The growth of structure can of course be
theoretically predicted when either one of the assumptions above is relaxed,
but those predictions tend to be more detail-dependent, have functional or numerical forms that are
necessarily more complicated, and are also in the regimes where the comparison
to data is more difficult (see the discussion in the bulleted list just
below). Therefore, we will largely stick to studying the growth as given in
Eq.~(\ref{eq:growth}), with the assumptions given in the bullet points
above. Before proceeding, however, we wish to comment further about a few of these
assumptions and situations in which relaxing them may be relevant:

\begin{itemize}
  \item \textbf{Beyond general relativity:} Our assumption of Einstein's general
    relativity in deriving Eq.~(\ref{eq:growth}) deserves particular
    attention, as modified-gravity theories generically lead to the
    scale dependence of $\delta(t)$. Observing this scale dependence, and
    ruling out the systematic errors as the cause, would present striking
    evidence for the presence of modified gravity. There has been growing
    interest in trying to constrain the scale-dependent predictions of the
    growth of structure with current and future surveys
    \cite{Zhao:2008bn,Zhao:2009fn,Daniel:2010yt,Zhao:2010dz,Song:2010fg,Silvestri:2013ne}.\\[-0.3cm]
  \item \textbf{Near-horizon scale:} Equation (\ref{eq:growth}) is modified
    on scales approaching the Hubble distance, $k\simeq H_0\simeq 10^{-3}\hmpcinv$. This is
    due to relativistic effects which can be precisely quantified
    \cite{Yoo:2009au,Yoo:2010ni,Challinor:2011bk,Jeong:2011as,Bonvin:2014owa,Tansella:2017rpi,Grimm:2020ays}.
    Detecting these relativistic effects would be a very interesting test of the standard
    cosmological model, is challenging given that they appear on very large
    scales, but may nevertheless be possible with forthcoming large-scale structure surveys
    \cite{Maartens:2012rh,Alonso:2015uua,Alonso:2015sfa,Fonseca:2015laa,Abramo:2017xnp,Barreira:2022sey}.\\[-0.3cm]
  \item \textbf{Beyond linear theory:} this assumption is the most ``ready''
    to be relaxed, as a large portion of current observations lies in the
    quasi-linear or fully nonlinear regime. In addition to pure
      gravitational nonlinearities which can be reasonably accurately modeled
      using a combination of numerical and analytic tools, these scales are
      also affected by effects of baryons which affect the clustering on
      scales of a few megaparsecs and below.
    Current surveys (e.g.\ KiDS
    \cite{KiDS:2020suj}, DES \cite{DES:2022ccp}, and HSC \cite{HSC:2018mrq})
    explicitly throw out scales that are strongly affected by baryons, but even then some
    information comes from quasi-linear regime where Eq.~(\ref{eq:growth})
    does not hold. This puts premium on our ability to model the growth of
    structure in this regime with a combination of numerical and analytic tools. Such
    predictions are available in $\Lambda$CDM (cosmological model with vacuum
    energy and cold dark matter) 
    \cite{Hamilton:1991es,Peacock:1996ci,Smith:2002dz,Heitmann:2009cu,Takahashi:2012em,Heitmann:2013bra,Mead:2015yca,Garrison:2017ssz,Bird:2018all,DeRose:2018xdj,Mead:2020vgs},
    but more recent work has extended this to dark energy models with the
    equation of state parameterized by $(w_0, w_a)$ \cite{Linder:2003dr,Francis:2007qa,Casarini:2016ysv,Lawrence:2017ost,Euclid:2020rfv},
    as well as modified-gravity models
    \cite{Stabenau:2006td,Laszlo:2007td,Oyaizu:2008tb,Baldi:2008ay,Koyama:2009me,Schmidt:2009sg,Chan:2009ew,Zhao:2010qy,Li:2013tda,Barreira:2014kra,Winther:2015wla,Bose:2016wms}.
\end{itemize}


\subsection{Growth in simple cosmological models}

The growth equation (\ref{eq:growth}) can of course be solved numerically for
an arbitrary cosmological model --- the input required is the scaling of the
Hubble parameter with time, $H(t)$, and that of the matter density,
$\rho_{\rm M}(t)$. However, it is instructive to derive scalings in single-component
universes, where the expansion is dominated by a single component --- matter,
radiation, or dark energy.

Let us start with the radiation-dominated case. Then $a\propto t^{1/2}$, so
that $H(t)\equiv \dot{a}/a\propto 1/(2t)$. Also, note that the last term in
Eq.~(\ref{eq:growth}) is negligible because the Hubble parameter is dominated
by the radiation and not matter density, so that $4\pi G\rho_{\rm M}\ll H^2$. Therefore,
we need to solve the equation
$\ddot\delta + 2H\dot\delta=0$,
with $H=1/(2t)$. Its solution is 
\begin{equation}
\delta(t) = A_1 + A_2\ln t\qquad (\mbox{radiation dominated}),
\end{equation}
where $A_1$ and $A_2$ are some constants.  
Thus, in the radiation-dominated regime, the fluctuations grow only very
slowly --- logarithmically with time.

In the flat, matter-dominated (Einstein-de Sitter) case, $a\propto t^{2/3}$ so
that $H(t)\equiv \dot{a}/a\propto 2/(3t)$, while the Hubble parameter is
dominated by matter density, so that $4\pi G\rho_{\rm M}= (3/2)H^2$. Let us assume
that $\delta(t)\propto t^n$; then the growth equation simplifies to $n(n-1) +
{4\over 3}n - {2\over 3} = 0$. Its solutions are easy to obtain: $n=+2/3$ and
$-1$. Hence
\begin{equation}
\delta(t) = B_1t^{2/3} + B_2t^{-1}\qquad (\mbox{matter dominated}),
\end{equation}
where $B_1$ and $B_2$ are some constants.  Since $a(t)\propto t^{2/3}$ in the
matter-dominated era, the growing mode of the perturbations grows
proportionally to the scale factor, $\delta(t) \propto a(t)\propto
t^{2/3}$. This scaling is of the utmost importance, as the universe spends of
order 10 billion years in the matter-dominated era --- from the
matter-radiation equality 50,000 years after the Big Bang, to the onset of
dark energy a few billion years ago. During that time, structures in the
universe grow appreciably, all thanks to the $\delta\propto a$ scaling.

Finally, in the dark-energy-dominated era, which will presumably take place in
the future when dark energy dominates completely, the scale factor grows
exponentially\footnote{The same happens during inflation, when the universe is
  completely vacuum-energy dominated.}, $a\propto e^{Ht}$, so that $H(t)\equiv
H_\Lambda={\rm const}$. [Here we are representing dark energy by a specific
  model, that of the cosmological constant $\Lambda$, which mathematically
  represents vacuum energy which has a constant energy density across cosmic
  time.] Also, note that the last term in Eq.~(\ref{eq:growth}) is negligible
since the matter density is negligible relative to vacuum energy in
$H$. Therefore, we need to solve the equation $\ddot\delta +
2H_\Lambda\dot\delta=0$ whose solution is
\begin{equation}
\delta(t) = C_1 + C_2 e^{-2H_\Lambda t}\simeq {\rm const}\qquad ({\rm Lambda\,\, dominated}),
\end{equation}
where $C_1$ and $C_2$ are some constants, and where the exponentially decaying
term becomes negligible quickly.  Therefore, density perturbations do not grow
at all in a Lambda-dominated universe.  We are all witnesses to this effect
today, as our universe with 70\% dark-energy and 30\% matter displays severely
suppressed structure formation (relative to the matter-only scenario),
something that is readily observed in the cosmological data that probe
$z\lesssim 1$ using tests that we describe in Sec.~\ref{sec:obs}.

Summarizing, the temporal evolution of linear density perturbations in simple,
single-component cosmological models is:
\begin{equation}
  \delta(a)\propto  
  \left \{  \begin{array}{cl}
    t^{2/3}\propto a & \mbox{(matter dominated)} \\[0.3cm]
    \ln (t) \propto \ln (a) & \mbox{(radiation dominated)}\\[0.35cm]
    {\rm const}  &  \mbox{(Lambda dominated)},
\end{array} \right .
\end{equation}
where $a$ is the scale factor.

\subsection{Dimensionless linear growth function}

It is useful to cast Eq.~(\ref{eq:growth}) in a dimensionless form. Let us
introduce the \textit{linear growth function}
\begin{equation}
D(a) = \frac{\delta(a)}{\delta(1)},
\end{equation}
where $D$ at the present time is unity, $D(1)=1$.
\begin{figure}[!t]
\begin{center}
\includegraphics[scale=0.6]{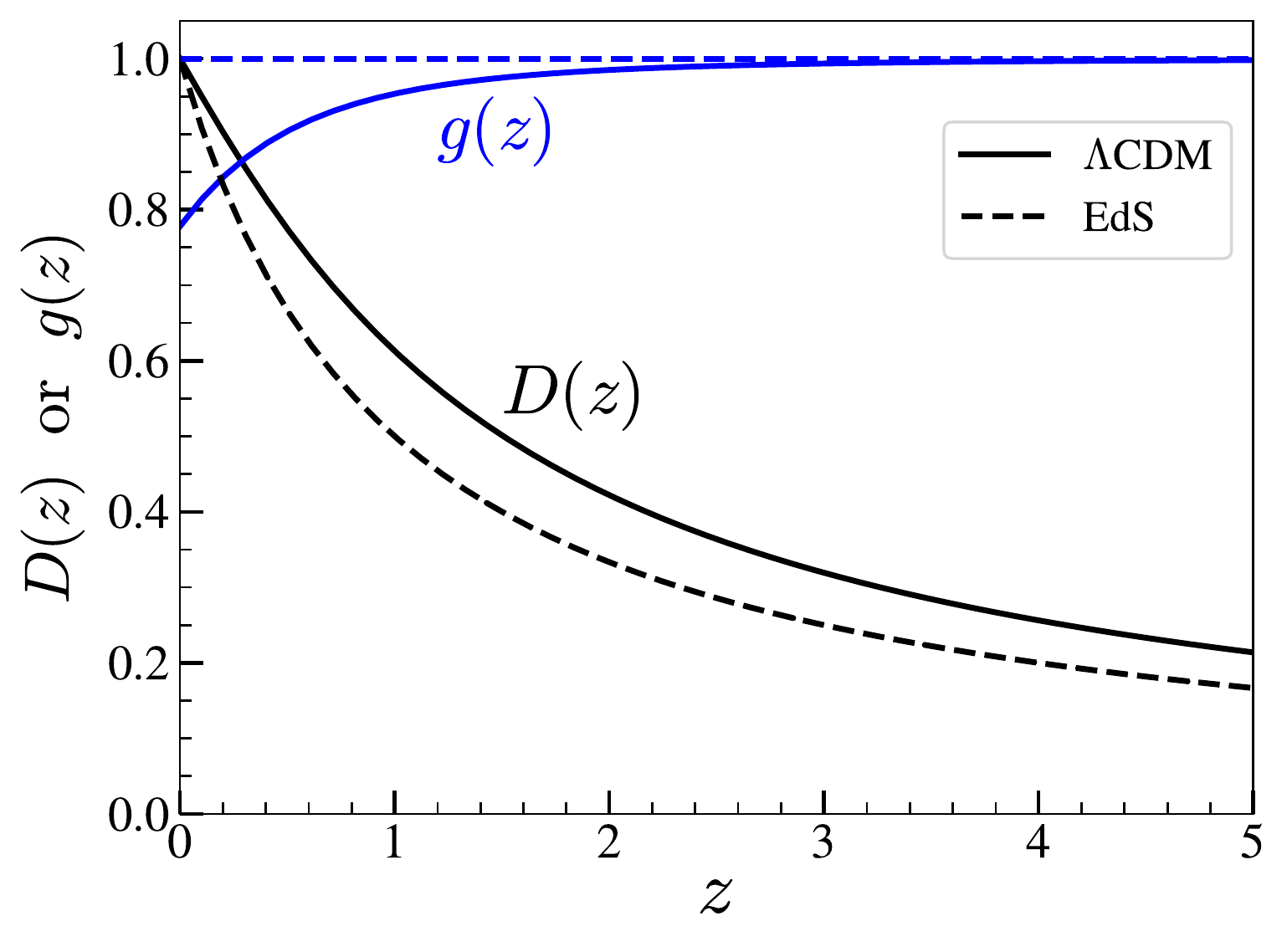}
\end{center}
\caption{Illustration of the linear growth function $D(z)$ and the growth
  suppression factor $g(z)$, as a function of redshift $z$, in the standard
  flat $\Lambda$CDM cosmological model. For comparison, we also show the
  results in the flat model with dark matter only ($\om=1$), the
  Einstein-de Sitter (EdS) model. Either $D(z)$ or $g(z)$ fully describes the
  linear growth of fluctuations; see text for details. }
\label{fig:linear_growth}
\end{figure}
The original equation determining the linear growth of structure,
Eq.~(\ref{eq:growth}), can now be set in a dimensionless form as
\begin{equation}
2\frac{d^2g}{d\ln a^2} +\left[5 - 3w(a)\Omega_{\rm DE}(a)\right]
\frac{dg}{d\ln a}+ 3\left[1-w(a)\right]\Omega_{\rm DE}(a)g=0,
\label{eq:g_suppression}
\end{equation}
\noindent
where $g\equiv g(a)$ is the \textit{growth suppression factor} --- that is, growth
relative to that in Einstein-de Sitter ($\om=1$) universe. The linear growth suppression factor
$g$ is related to the growth factor $D$ implicitly via
\begin{equation}
  D(a)\equiv {ag(a)\over g(1)}.
  \label{eq:D_g_rel}
\end{equation}
Note that the linear growth function depends on the Hubble parameter $H(a)$,
as well as the matter density $\om$. Thus the linear growth function
depends on basic cosmological parameters; in the flat $\Lambda$CDM model for
example, $D=D(a, \om)$, and same for $g$. In particular, the growth
function does
not depend on wavenumber $k$. [This is no longer true in modified-gravity
  models as explained above, but also in models with massive neutrinos, which
  introduce a mild scale-dependence to the linear growth function
  \cite{Lesgourgues:2006nd}.]

Figure \ref{fig:linear_growth} shows the temporal evolution of $D(z)$ and
$g(z)$ in the standard cosmological $\Lambda$CDM model with
$\om=1-\Omega_\Lambda=0.3$ (in what follows, we will frequently switch
between redshift $z$ and scale factor $a=1/(1+z)$). For comparison, we also
show the results in the flat model with dark matter only ($\om=1$), the
Einstein-de Sitter (EdS) model.  In $\Lambda$CDM, the onset of dark energy at
late times ($z\lesssim 1$) causes these functions to deviate from their EdS
behavior ($D(a)=a$ or $D(z)=1/(1+z)$, and $g(z)=1$). The late-time decrease in
either $g(z)$ or $D(z)$ fully specifies the suppression in the growth of
structure (in linear theory) due to dark energy.  In our fiducial flat
$\Lambda$CDM cosmological model with $\om=1-\Omega_{\rm DE}=0.3$, the
present-day value of the suppression factor is $g(z=0)\simeq 0.78$.

Linear growth function enters the linear matter power spectrum in a
straightforward way.  The matter power spectrum, $P(k)$, quantifies the amount of
structure --- the power --- at each wavenumber $k$. The power spectrum is defined in terms of
the two-point correlation function of Fourier-space overdensity $\delta_\bfk$ as
\begin{equation}
\langle \delta_{\bfk}\,\delta_{\bfk'}^*\rangle = (2\pi)^3\,
\delta^{(3)}(\bfk-\bfk')\,P(k, a)
\label{eq:Pk}
\end{equation}
where the angular brackets denote ensemble average, and $\delta^{(3)}$ is the
Dirac delta function. Note that the power spectrum depends only on the magnitude of the wavenumber,
$k=\lvert\bfk\vert$, due to the assumptions of homogeneity and isotropy.
Because $\delta_\bfk(a)\propto D(a)$ by definition, the dependence of the
linear matter power spectrum on the  linear  growth is simple enough,
\begin{equation}
  P(k, a) \propto D(a)^2 \propto [ag(a)]^2,
  \label{eq:Pk_growth}
\end{equation}
or, entirely equivalently, $P(k, z)\propto D(z)^2$. Therefore, a simple
observation of how galaxy clustering scales with redshift is sensitive to the
growth of structure. This is, however, only one way to probe the growth, as
cosmological observations are sensitive to its different aspects, only some of
which are captured in Eq.~(\ref{eq:Pk_growth}). We now cover this in more detail.
  
\section{Connection to observations}\label{sec:obs}

There are several ways in which one can measure the growth of structure
through cosmic time.  Measuring galaxy clustering in redshift effectively
measures $P(k, z)$ in (typically) several redshift bins, thus probing
growth. Additionally, one can map out the cosmic shear pattern, measuring
correlations of shapes of distant galaxies which is sensitive to the
distribution of mass along the line of sight and thus the growth of
structure. Cross-correlating galaxy clustering and cosmic-shear signal is
another method that can provide additional information on growth. A rather
different approach entails measuring the space density of clusters of
galaxies; their density as a function of mass and redshift --- the mass
function --- depends on the linear growth function $D(z)$.  Finally, one can
measure the correlation of galaxy velocities as it too is sensitive to
growth. We now discuss these cosmological probes, and how growth enters them.

\subsection{Galaxy clustering }\label{sec:gg}

Because galaxy clustering is one of the principal
cosmological probes, its measurements provide the principal way to isolate and
constrain the growth of cosmic structure. Measurements of galaxy clustering
constrain the power spectrum $P(k, z)$ or, equivalently, the two-point
correlation function $\xi(r, z)$, over a range of scales and in several
redshift bins. It is the dependence on redshift $z$ that informs us about the
temporal growth of cosmic structure.

We now quantify the dependence of clustering on growth.  The linear
($\lvert\delta\rvert\ll 1$) dimensionless matter power spectrum $\Delta^2(k, z)$ can
be expressed in terms of $P(k)$ as
\begin{equation}
  \begin{aligned}
  \Delta^2(k, z)&\equiv \frac{k^3 P(k, z)}{2\pi^2}\\[0.2cm]
  &=A_s\,{4\over 25}
        {1\over \om^2}\left ({k\over k_{\rm piv}}\right )^{n_s-1}
        \left ({k\over H_0}\right )^4 \,\left (\frac{g(z)}{1+z}\right )^2\, T^2(k),
        \label{eq:Deltasq}
        \end{aligned}
\end{equation}
where we expressed the temporal dependence in terms of redshift $z$.  Here
$A_s$ is the normalization of the power spectrum (for the fiducial cosmology,
$A_s\simeq 2.1\times 10^{-9}$), $n_s$ is the spectral index, $k_{\rm piv}$ is
the ``pivot'' around\footnote{Typically tajen to be $k_{\rm
    piv}=0.05\Mpc^{-1}$, which is close to the wavenumber at which the
  primordial power is best constrained.} which $\Delta^2(k)$ varies as a power
law in wavenumber, the combination $ag(a)\equiv g(z)/(1+z)\propto D(z)$ determines
the linear growth of perturbations, and $T(k)$ is the linear transfer function
which mainly encodes the change in the shape of the power spectrum around the scale
corresponding to horizon size at matter-radiation equality.

Dependence of the linear matter power spectrum on the (linear) growth in
Eq.~(\ref{eq:Deltasq}) is simple enough, $P(k, z)\propto g(z)^2\propto D(z)^2$. However, in
practice there are two complications:
\begin{itemize}
  \item First, we can measure the power spectrum of galaxies and not
    dark-matter particles. Traditionally, the relationship between the
      galaxy overdensity $(\delta \rho)_g/\rho_g$ and the matter overdensity
      $\delta\rho/\rho$ is given by the so-called galaxy bias term, $b\equiv
      [(\delta \rho)_g/\rho_g] / [(\delta \rho)/\rho]$. The relationship
        between the galaxy power spectrum and the matter power spectrum is
        consequently
\begin{equation}
  P_{gg}(k, z) = b^2(k, z) P(k, z) 
  \label{eq:P_gg}
\end{equation}
where $b(k, z)$ is galaxy bias which depends on the wavenumber in a way that
may be difficult to predict theoretically. Further, galaxy bias depends on the
galaxy type and, worse, on the galaxy formation history (often termed
``assembly bias''), and thus typically needs to be measured directly from the
data. Even on linear scales, where galaxy bias is expected to be
scale-independent (i.e.\ constant in wavenumber $k$), its time dependence is a
priori unknown. This time dependence of the bias is unfortunately degenerate
with that of the growth of structure, and breaking this degeneracy requires
either independent prior information on the bias or else combination of galaxy
clustering with other cosmological measurements.\footnote{There is no
  way to break this degeneracy between the time-dependence of bias and growth
  with galaxy clustering alone. However, as we discuss in
  Sec.~\ref{sec:shear}, one can use weak gravitational lensing (which
  altogether avoids galaxy bias), as well as galaxy-galaxy lensing
  (proportional to bias, rather than bias squared), to break this degeneracy
  and isolate the growth of structure. Alternatively, simultaneous
  measurements of the power spectrum and other statistics of galaxy clustering
  (e.g.\ the three-point correlation function, or the bispectrum) can help
  separately constrain galaxy bias and the growth of structure.}
\item The second complication is that typical clustering measurements are made
  partly in the quasi-linear and non-linear regime, corresponding roughly to
  scales $r\lesssim 10\hinvmpc$ (or $k\gtrsim 0.1\hmpcinv$) at
  $z=0$. Nonlinear corrections to the matter power spectrum also depend on
  scale, while additional scale dependence in galaxy clustering is
    brought about by baryonic effects, as well as the dependence of galaxy
    bias $b(k,z)$ on scale.  The resulting effects on the galaxy power
    spectrum are not theoretically tractable and must be calibrated with N-body
    simulations --- ideally hydrodynamical simulations which contain both the
    baryon and the dark matter particles.
\end{itemize}

Traditionally, cosmological constraints from galaxy clustering have been
quoted in terms of constraints on the amplitude of mass fluctuations
$\sigma(z, R)$. This quantity effectively averages the matter power spectrum
evaluated at some redshift $z$ over a spherical region of comoving radius
$R$. Mathematically,
\begin{equation} 
\sigma^2(z, R) = \int_0^\infty \Delta^2(k, z)
\left ({3j_1(kR)\over kR}\right )^2 d\ln k,
\label{eq:sigma}
\end{equation}
where $\Delta^2(z, k)$ is the dimensionless power spectrum from
Eq.~(\ref{eq:Deltasq}).  The galaxy-clustering constraints are typically
converted to those on the amplitude of mass fluctuations evaluated at $z=0$
and $R = 8\hinvmpc$ --- hence we often make use of the quantity
$\sigma_8\equiv \sigma(z=0, R = 8\hinvmpc)$.

Galaxy clustering is a mature cosmological probe, as its first measurements
date to 1960s and 70s. It is also reasonably easily accessible --- one only
needs to measure the galaxy positions in order to calculate the correlation
function. Ideally galaxy redshifts are available, in which case the full 3D power
spectrum can be computed, otherwise one needs to determine the redshifts
approximately using photometric information from galaxies (for a review of
photometric redshifts, see \cite{Newman:2022rbn}).  However, the comparison of
galaxy clustering with theoretical predictions is seriously challenged by the
presence of galaxy bias, as we discussed above. This is where weak
gravitational lensing comes in as a very powerful complement; we discuss this
next.

\index{DES (Dark Energy Survey)}
\begin{figure}[t]
\includegraphics[width=\textwidth]{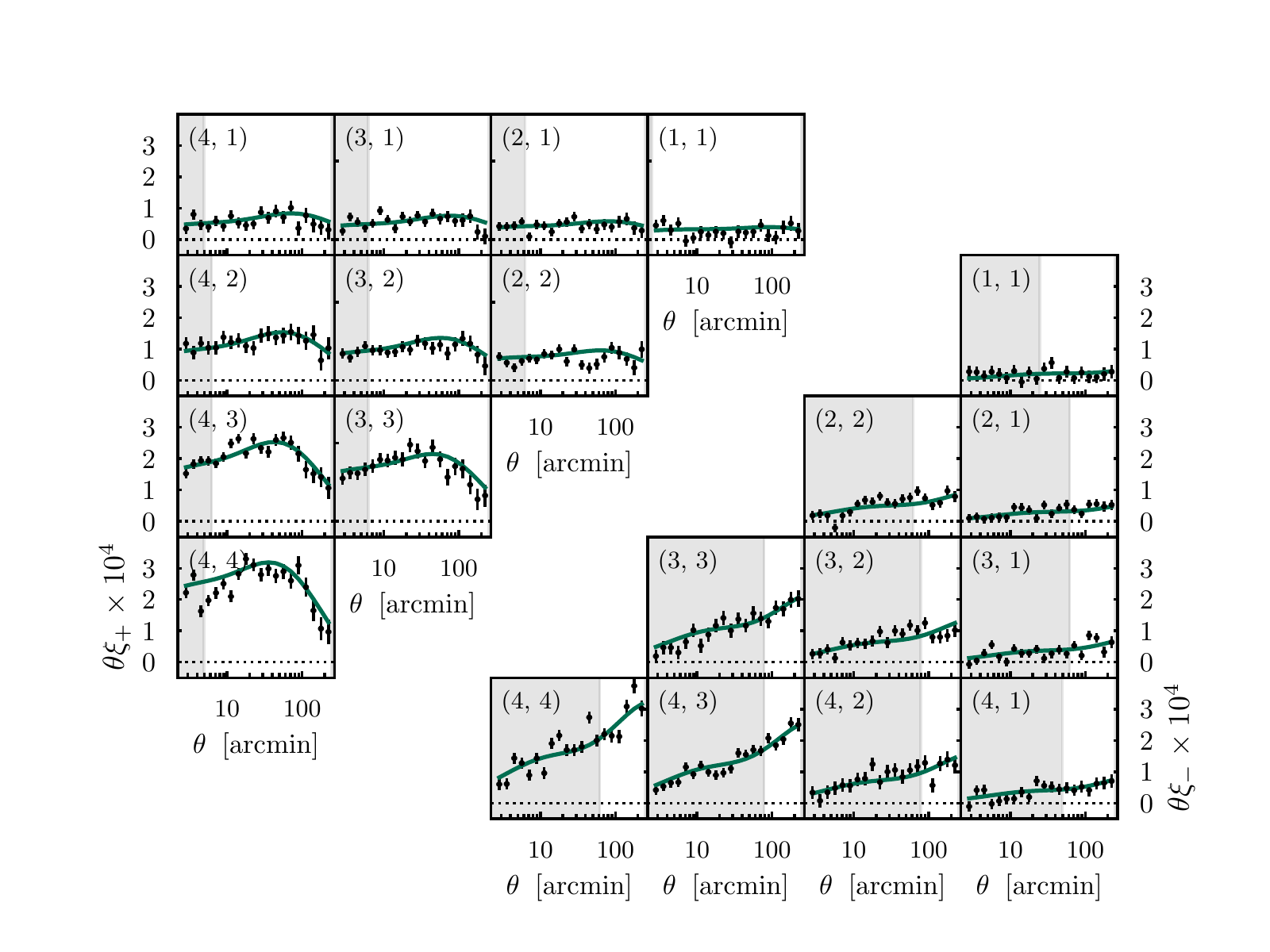}
\caption{Tomographic shear measurements of the 2-point correlation function of
  shear in real space, $\xi^{ij}_{\pm}(\theta)$ from the Year-3 data analysis
  of the Dark Energy Survey. The pair of numbers in $\xi$ (written in the top
  left corner of each panel) refers to the pair of redshift bins from which
  the measurement is made (e.g.\ ``2,4'' means spatial correlation of galaxies
  in bin 2 to those in bin 4). The plots above the diagonal show
  $\theta\xi^{ij}_{+}(\theta)$ as a function of $\theta$, while those below
  show $\theta\xi^{ij}_-(\theta)$. The curvy solid line shows the best-fit
  $\Lambda$CDM model. The grey regions show scales \textit{not}
  used in the final parameter constraints due to concerns about modeling of
  the nonlinear clustering at these smaller scales.  Adopted from DES Y3
  shear analysis \cite{DES:2021vln}.}
\label{fig:DES_WL}
\end{figure}

\subsection{Cosmic shear}\label{sec:shear}

Weak gravitational lensing --- subtle distortions in galaxy shapes due to the
intervening large-scale structure --- is a very powerful probe of the growth
of cosmic structure. Weak lensing is sensitive to the presence and distribution
of mass along the line of sight, between the observer and source
galaxies. This method, proposed in the 1960s and first detected in the year 2000, is
now a standard-bearer for the probes of large-scale structure. Weak lensing
also goes under the name \textit{cosmic shear} as one statistically measures the
amount of shearing of the observed shape of galaxies caused by photon
deflections, which in turn informs about the projected mass along the line of sight.
See Refs.~\cite{Bartelmann:1999yn,Hoekstra:2008db} for reviews of weak lensing
/ cosmic shear.

The principal feature of cosmic shear is the absence of galaxy bias. Even
though measuring galaxy shapes (for cosmic shear) is more challenging than
measuring galaxy positions (for galaxy clustering), this absence of possible
degeneracies between the bias and the growth of structure makes cosmic shear a
premier cosmological probe.

A key quantity in cosmic shear is the convergence $\kappa$, which is defined
at every point on the sky and is proportional to the projected matter density
between the observer and the source galaxy. Specifically, convergence for a
single lens and a single source galaxy is given
by
\begin{equation}
\kappa
=
  \frac{d_{\rm L} d_{\rm LS}}{d_{\rm S}}\int_0^{\chi_{\rm S}} \nabla^2 \Phi\,d\chi,
\label{eq:convergence}
\end{equation}
where $d_{\rm L}$, $d_{\rm S}$, and $d_{\rm LS}$ are respectively the distance
from the observer to the lens and to the source, and the distance between the
lens and the source. Here, $d$ is the angular diameter distance which is
  related to the comoving distance $r$  via
  $d(z)=r(z)/(1+z)$, while $\chi$  is the coordinate distance and is related to the
  comoving distance via standard relations involving a sine or a sinh
  \cite{Huterer_book}; in a flat universe, $r=\chi$. Further, $\Phi$ is the three-dimensional gravitational
potential that gets integrated along the line of sight in the above equation.
A closely related quantity is shear $\gamma$ which, along with the
convergence, makes up elements of a $2\times 2$ matrix whose components define
the distortion of an image at any point on the sky.

Transforming the convergence to harmonic space and assuming statistical
isotropy, one obtains the convergence power spectrum $P^{\kappa\kappa}$,
defined as (e.g.\ \cite{Huterer_book})
\begin{equation}
\langle \kappa_{\ell m}\kappa_{\ell'm'}\rangle =
\delta_{\ell \ell'}\, \delta_{mm'}  \,P^{\kappa\kappa}(\ell),
\label{eq:Pkappa_vs_klm_klm}
\end{equation}
where the multipole $\ell$ correspond to an angle $\theta\simeq
180^\circ/\ell$.  The convergence power spectrum is identical to the shear
power spectrum in the limit of weak distortions (i.e.\ in the weak-lensing
limit); $P^{\gamma\gamma}(\ell)\simeq P^{\kappa\kappa}(\ell)$.  For a Gaussian
field, either $P^{\gamma\gamma}$ or $P^{\kappa \kappa}$ would contain all
information. Beacuse the weak lensing field is nongaussian on small scales,
higher-order correlations contain additional information and may be useful to
exploit.

The convergence (or shear) power spectrum is related to theory in a
straightforward way: it is a projection along the line of sight of the
three-dimensional matter power spectrum $P(k)$. Here we also assume weak
lensing \textit{tomography} \cite{Hu:2002rm} --- slicing of the shear signal
in redshift bins --- which enables extraction of additional information from
the weak lensing shear, as it makes use of the radial information which is
crucial for growth. Consider correlating shears in some redshift bin $i$ to
those in redshift bin $j$. Tomographic cross-power spectrum for these two
redshift bins, at a given multipole $\ell$, can be related to theory by
starting with Eqs.~(\ref{eq:convergence}) and (\ref{eq:Pkappa_vs_klm_klm});
the result is
\begin{equation}
P_{ij}^{\kappa\kappa}(\ell)\simeq P_{ij}^{\gamma\gamma}(\ell) = 
\int_0^{\infty} dz \,{W_i(z)\,W_j(z) \over r(z)^2\,H(z)}\,
P\! \left ({\ell\over r(z)}, z\right ). 
\label{eq:Pkappa_ij}
\end{equation} 
Here   the  weights $W_i$ are given by 
\begin{equation}
  W_i(\chi)\equiv \frac{3}{2}\,\om H_0^2\,q_i(\chi)\, (1+z),
  \label{Wchi_wl_tomo}
\end{equation}
where
\begin{equation}
  q_i(\chi)\equiv r(\chi)\int_{\chi}^{\infty} d\chi_{\rm S} n_i(\chi_{\rm S}) \frac{r(\chi_{\rm S}-\chi)}{r(\chi_{\rm S})},
\end{equation}  
and $n_i$ is the normalized ($\int n(z)dz=1$) comoving density of galaxies if
the coordinate distance $\chi_{\rm S}$ falls in the distance range bounded by the
$i$th redshift bin, and zero otherwise. Further, $r(z)$ is the comoving
distance and $H(z)$ is the Hubble parameter. Note that the
weights $W_i$ are purely geometrical, and do not contain any growth
information. The key takeaway from Eq.~(\ref{eq:Pkappa_ij}) is that
weak-lensing convergence power spectrum is proportional to the matter power
spectrum $P(k)$, and hence (on linear scales) to the growth squared, $D(z)^2$,
but without the complicating presence of galaxy bias.

Weak lensing therefore offers particularly good prospects for constraining the
growth of structure. With $N$ tomographic bins, one has a total of $N(N+1)/2$
tomographic bin pairs, each of which contains cosmological information. For
typical surveys $N\simeq 5$-$10$, enabling in principle the constraints on the
growth of structure $D(z)$ with relatively fine temporal resolution.

Modern measurements of the tomographic power spectrum of weak lensing
  shear, adopted from  the Dark Energy Survey (DES; \cite{DES:2021vln}),
are shown in Fig.~\ref{fig:DES_WL}. Note that the figure shows the real-space
shear correlation functions $\xi^{ij}_\pm(\theta)$ (multiplied by $\theta$ for
readability) for any two tomographic bins $i$ and $j$. These real-space
correlation functions are just harmonic transforms of the convergence power
spectrum $P_{ij}^\kappa(\ell)$, and are used because the weak-lensing measurements
are often performed in real space (in $\theta$), rather than in multipole
space (in $\ell$). The real-space correlation functions are obtained from the convergence power
spectrum via
\begin{equation}
  \xi^{ij}_\pm(\theta) = \frac{1}{2\pi}\int_0^\infty
  P^\kappa_{ij}(\ell)J_{0,4}(\ell\theta)\ell d\ell,
\end{equation}
where $J_{0,4}(x)$ is the Bessel function of zeroth and fourth order,
respectively. Note the impressive agreement between the shear measurements and
the best-fit $\Lambda$CDM model in Fig.~\ref{fig:DES_WL}, extending even to
the scales that were conservatively not used in the cosmological analysis
(shaded in gray).

There are also important systematic errors that enter weak-lensing
measurements. Notably, shear measurements are in general challenging, and are
impacted by smearing of galaxy shapes by atmospheric distortions or other
instrumental artifacts. Intrinsic alignments between source galaxies \cite{Catelan:2000vm,Hirata:2004gc}
complicate theoretical predictions and need to be explicitly modeled. Finally,
photometric redshifts of source (and lens) galaxies have complex statistical
properties that need to be corrected for where possible, then explicitly
modeled in the analysis. For a review of these and other issues, see
\cite{Mandelbaum:2017jpr}.

\subsection{Galaxy-shear cross-correlations and 3$\times$2 analysis}  \label{sec:3x2}

In addition to measuring shapes of distant galaxies across the sky and
correlating them, there are other ways to leverage weak lensing observations
to learn about the growth of structure in the universe.  One option is to measure the
correlation of the \textit{shears} of background galaxies with the
\textit{positions} of the foreground galaxies.  This correlation is known
under the name \textit{galaxy-galaxy lensing}, and effectively measures the
lensing efficiency of foreground galaxies.  The method should perhaps be
called galaxy-galax\textit{ies} lensing, as it correlates the position of one
foreground galaxy with the shapes of a number of background galaxies that are near
it on the sky.  Another better name is \textit{galaxy-shear
  cross-correlation}, as galaxy positions are correlated with (other
galaxies') shear.

Modern analyses combine the galaxy clustering (galaxy-galaxy correlations),
cosmic shear (shear-shear correlations), and galaxy-galaxy lensing
(galaxy-shear correlations) into one unified analysis. This combination goes
under the informal name of ``3$\times$2'' analysis, named for three two-point
correlation functions.

If the actual observations were made in a single redshift bin and on a single
spatial scale, then they would
correspond to the data vector of three elements, $[gg, g\gamma,
  \gamma\gamma]$, where $g$ stands for galaxy position, and $\gamma$ for
cosmic shear. Because the observations span a number of redshifts and scales,
the resulting data vector can be much longer. For example, in the DES Y3
analysis \cite{DES:2021wwk}, the full data vector contains about 1000
elements, corresponding to measurements of $gg$, $g\gamma$, or $\gamma\gamma$,
evaluated at a number of spatial scales $k$ and in several redshift
bins. About half of these measurements were made on linear or quasi-linear
scales where theoretical modeling is accurate and were used to constrain
cosmological parameters.

Because a 3$\times$2 analysis efficiently combines observations of galaxy
clustering and gravitational lensing, it also serves as a major probe of the
growth of structure. The constraints on growth from a 3$\times$2-type analysis are not
trivial to isolate; we will discuss in Sec.~\ref{sec:grow-geom} how this can be
done. Instead, we typically quote constraints on the combination of the
amplitude of mass fluctuations and matter density that is well constrained by
galaxy clustering and weak lensing\footnote{Note that $S_8$ depends on
  redshift because $S_8(z)\propto\sigma_8(z)\propto D(z)$; recall
  Eq.~(\ref{eq:sigma}). Conventionally, the redshift-dependent part is taken
  out to quote constraints on $S_8$ at $z=0$.}
\begin{equation}
  S_8\equiv \sigma_8 \sqrt\frac{\om}{0.3}.
\end{equation}
Constraints on $S_8$ are then roughly interpreted as those on the growth of
cosmic structure. In Sec.~\ref{sec:constraints} below, we show a compilation of
current constraints on $S_8$ from current data.

\subsection{Counts of galaxy clusters}  \label{sec:clusters}

The abundance of clusters of galaxies as a function of their mass and redshift
is very sensitive to cosmic growth. Galaxy clusters are the observed baryonic
components of dark-matter \textit{halos}, so the abundance of clusters can
often be quantified via our theoretical understanding of the abundance of
halos.  The halo abundance formalism can be studied from first principles
using the excursion set formalism pioneered by Bond et al \cite{Bond:1990iw}
and reviewed by Zentner \cite{Zentner:2006vw}. In this formalism, formation of
a halo occurs when random walk of overdensity $\delta$, as a function of
radius of a sphere over which it is evaluated, crosses some critical
threshold. This threshold is quoted in terms of the peak height $\nu (M)\equiv
\delta_c/\sigma (M)$, where $\delta_c\simeq 1.686$ is the critical overdensity for
collapse and $\sigma$ is the amplitude of mass fluctuations defined in
Eq.~(\ref{eq:sigma}). Because halo formation is exponentially sensitive to the
peak height (in the simple spherical-top-hat-overdensity model), and
$\nu (z)\propto 1/\sigma(z)\propto 1/D(z)$, the sensitivity of the halo abundance to
the growth of structure is very strong.

Because galaxy clusters are the most massive collapsed objects in the universe, they
are described by the density field in the not-overly nonlinear regime. As a
consequence, their abundance can be modeled analytically, using arguments
first laid out by Press and Schechter \cite{Press:1973iz} that preceded those
of the more general (and aforementioned) excursion-set formalism, and with a rather
transparent dependency on the cosmological parameters
\cite{Allen:2011zs,Haiman:2000bw}. The quantity that describes the abundance
of galaxy clusters is the mass function, which is the space density at
arbitrary redshift and mass, $n(z, M)$, and is typically reported as the number per
unit log interval in mass, $dn/d\ln M$. The number of clusters with
mass above $M_{\rm min}$ and below redshift $z_{\rm max}$ in some volume of solid
angle $\Omega_{\rm sky}$ is then
\begin{equation}
  N(z<z_{\rm max}, M>M_{\rm min}) = \Omega_{\rm sky}\int_0^{z_{\rm max}} dz\int_{M_{\rm min}}^\infty \frac{dn}{d\ln M}
  \frac{r^2(z)}{H(z)}d\ln M,
\end{equation}
where $r^2(z)/H(z)=dV/(d\Omega dz)$ is the volume element, with $r$ the
comoving distance and $H$ the Hubble parameter. In practice, the number of
clusters is evaluated by connecting it explicitly to directly observable
quantities (as discussed near the end of this subsection), but the mass
function retains its central role in connecting observations with theory, and
its strong dependence on the growth of structure remains useful.

\begin{figure}[!t]
\begin{center}
\includegraphics[scale=0.37]{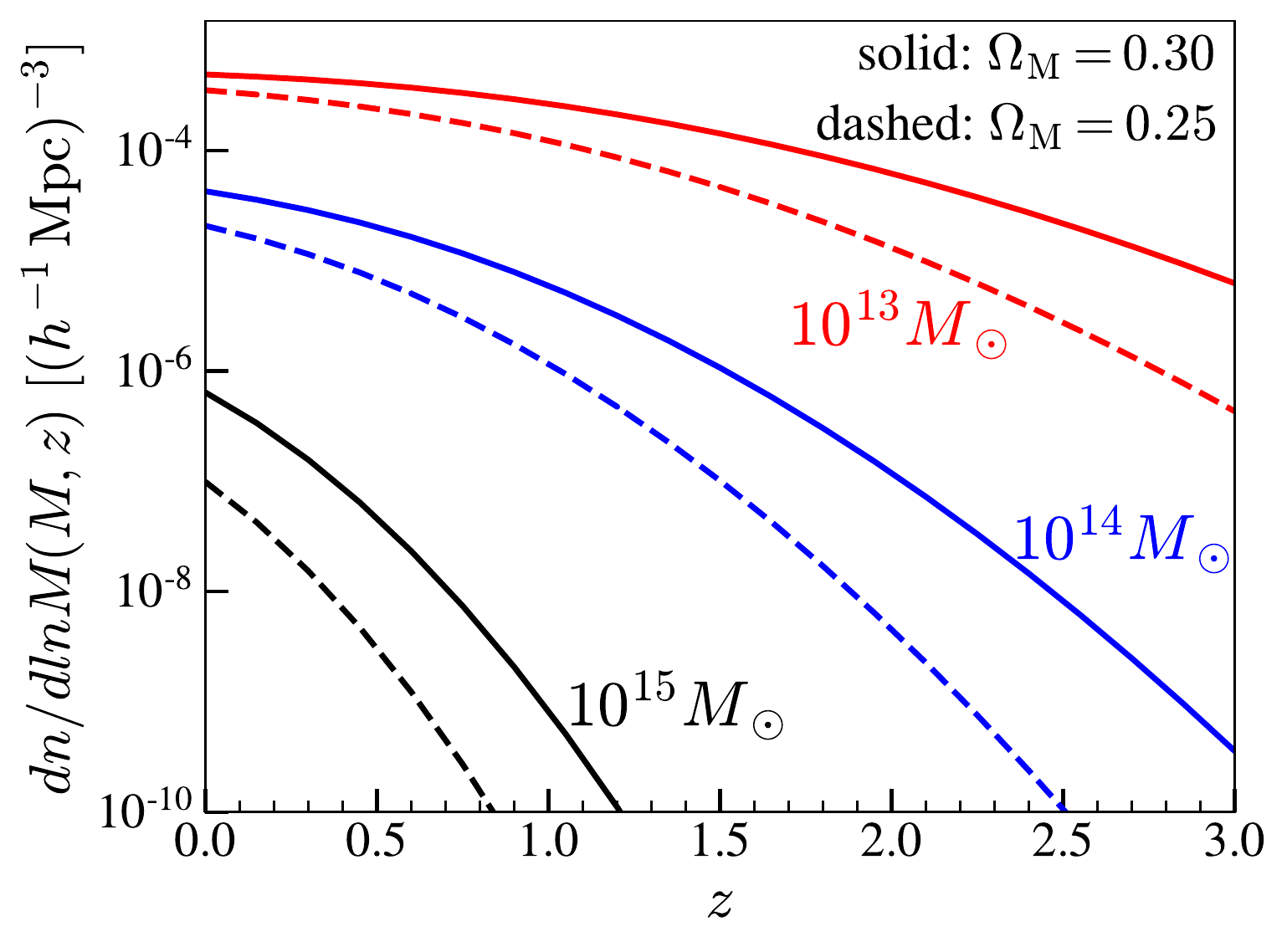}
\includegraphics[scale=0.37]{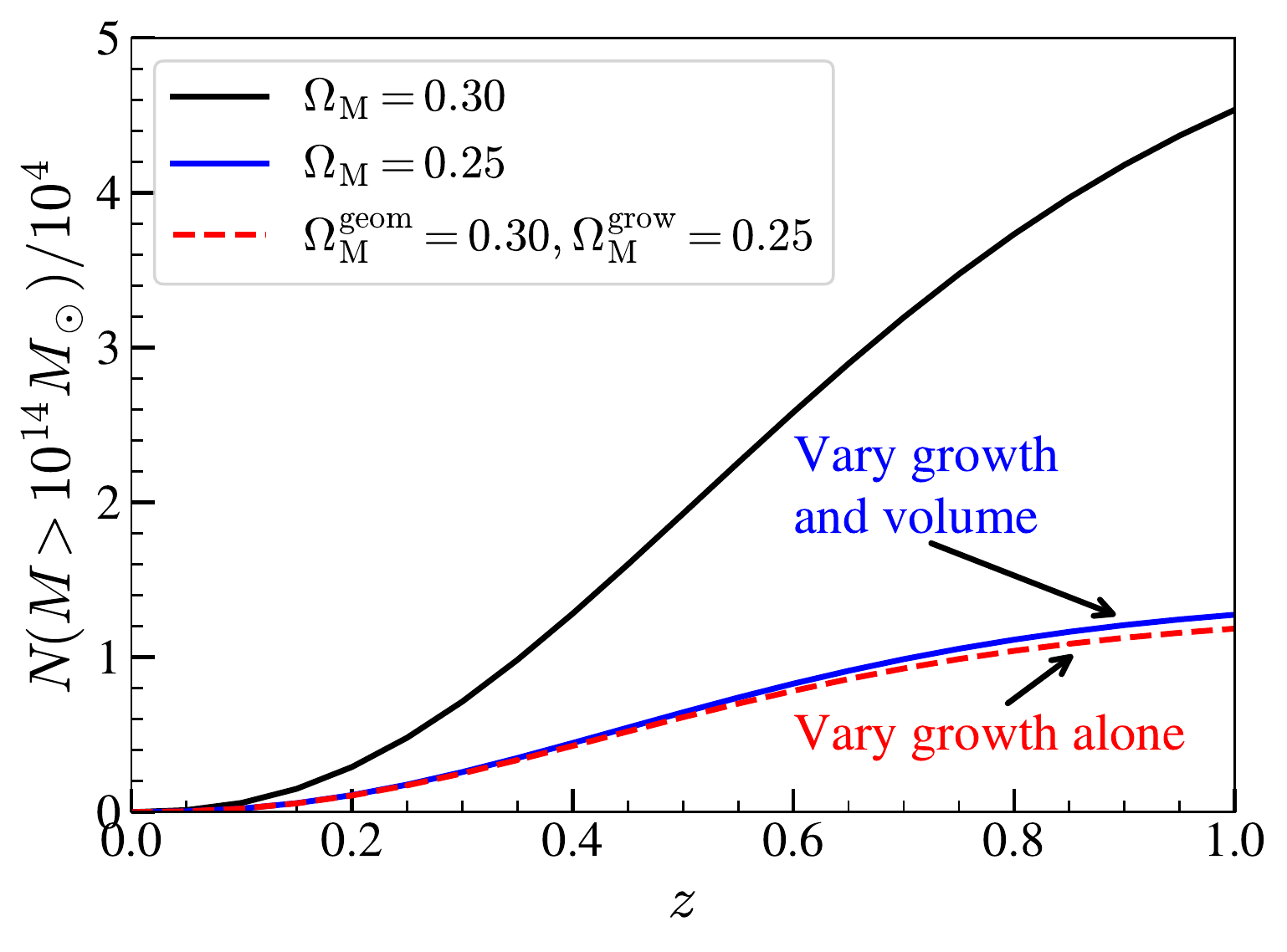}
\end{center}
\caption{Sensitivity of galaxy cluster counts on growth. The left panel shows
  the Tinker et al.~\cite{Tinker:2008ff} mass function, as a function of
  redshift, for $M=10^{13}, 10^{14}$, and $10^{15}\Msun$. The right panel
  shows the number of clusters expected above $M=10^{14}\Msun$ out to redshift
  $z$, assuming a survey covering 5000 sq.\ deg (note that the y-scale shows
  counts in units of $10^{4}$). The right panel shows that the sensitivity of
  number counts on cosmology comes largely through growth, since changing the
  growth alone (and keeping geometrical terms unchanged) accounts for most of
  the difference when moving to a new cosmological model --- here, a different
  value of $\om$.}
\label{fig:clusters}
\end{figure}

While the cutting-edge approach for modeling the mass function is to calibrate
it directly on simulations and interpolate in space of cosmological parameters
using a so-called emulator
\cite{McClintock:2018uyf,Nishimichi:2018etk,Bocquet:2020tes}, here we stick
with the semi-analytical mass function for illustrative purposes. We adopt the
Tinker \cite{Tinker:2008ff} mass function, which has been shown to be accurate
to $\sim5\%$ over a respectably wide range of mass and redshift. The Tinker mass
function is defined  as
$dM$ and $dz$ as
\begin{equation}
  \begin{aligned}
\frac{dn}{d\ln M} &=
f(\sigma) \frac{\rho_{\rm{M}, 0}}{M}\,
\frac{d\ln\sigma^{-1}}{d\ln M}  \\[0.2cm]
f(\sigma) &= A\left [\left (\frac{\sigma}{b}\right )^{-a}+1\right ]
e^{-c/\sigma^2},
\label{eq:MF}
  \end{aligned}
\end{equation}
where $\rho_{\rm{M}, 0}$ is the matter density today, $\sigma$ is the amplitude of
mass fluctuations, and $a, b$, and $c$ are redshift-dependent coefficients
calibrated from simulations and reported in Ref.~\cite{Tinker:2008ff}.
Moreover, $M$ refers to halo mass specifically defined as the mass in a
spherical region with 200 times the mean mass density in the universe.  Here,
the growth function dependence enters through the amplitude of mass
fluctuations $\sigma(M, z)\propto D(z)$ which is found in the exponential, as
predicted by the Press-Schechter (and, more generally, excursion-set) arguments.

In Figure \ref{fig:clusters} we show the sensitivity of the cluster counts on
the growth of structure. The left panel shows the Tinker mass function
vs.\ redshift for three values of mass ($M=10^{13}, 10^{14}$, and
$10^{15}\Msun$), and for two values of matter density $\om$. It shows
strong dependence of the expected number density of clusters on the matter
density, which in turn determines the growth of structure.  The right panel
shows the number of clusters expected above $M=10^{14}\Msun$ out to redshift
$z$, assuming a survey covering 5000 sq.\ deg (note that the y-scale shows
counts in units of $10^{4}$). We observe that the sensitivity of number counts
on cosmology comes largely through growth, rather than the geometrical volume
factor.  This explicitly illustrates the fact that cluster counts are very
sensitive probes of growth
\cite{Bahcall:1998ur,Haiman:2000bw,Mullis:2005hp,Brodwin:2010ig,Holz:2010ck,Kravtsov:2012zs,Mortonson:2010mj,Jee:2013gey}

Because cluster masses are not directly measured, modern measurements of the
abundance of clusters are typically compared to theory not in terms of their
masses, but rather other intermediate, more readily observable quantities
called mass ``proxies''. One such mass proxy is ``richness'', defined as the
number of galaxies that reside in a cluster. Other mass proxies include X-ray
flux or weak-lensing signal measured toward clusters; all of these proxies
correlate with cluster mass. Conversion from noisy measurements of mass
proxies to actual cluster masses introduces both statistical and systematic
errors, and controlling and quantifying these errors --- especially the
systematics ---  is the principal challenge for cluster cosmology. If the systematics can
be controlled and understood, then the prospects for constraining the growth
of structure via cluster abundance with ongoing or upcoming
wide-field cosmological surveys such as eROSITA, LSST, and Euclid and Nancy Roman
Space Telescopes are very good.

\subsection{Cosmic velocities}\label{pecvel}

Cosmic velocities are also sensitive to the growth of structure. In linear
theory, the velocity $\bfv$ is directly related to overdensity $\delta$ via the
continuity equation which reads

\begin{equation}
  \bfv = \frac{i\bfk}{k^2}\frac{D'}{D}\delta = \frac{i\bfk}{k^2}afH \delta,
\end{equation}
where the prime denotes a derivative with respect to conformal time $\eta$,
and the second equality follows because $' = d/d\eta = a(d/dt) = aH(d/d\ln
a)$.  Here we have defined the growth rate $f$ as
\begin{equation}
  f\equiv \frac{d\ln D}{d\ln a}.
\end{equation}
Then the velocity power spectrum can be related to the matter power spectrum
in linear theory
 via 
\begin{equation}
  P_{vv}(k, a) = \left [\frac{af(a)H(a)}{k}\right ]^2P(k, a).
 \label{eq:P_vv}
\end{equation}
Note a key feature: the velocity power spectrum not only scales as the
matter power spectrum (and hence the usual growth term $D(a)$ squared), but is
further proportional to the square of the growth rate $f(a)$. Because the
growth rate is sensitive to both the standard cosmological parameters
($\om$ and $w$, for example) and modified gravity, this latter dependence
makes the cosmic velocities particularly well suited to cosmological tests that rely on the
growth of structure \cite{Song:2008qt}.

\begin{figure}[!t]
\begin{center}
\includegraphics[scale=0.25]{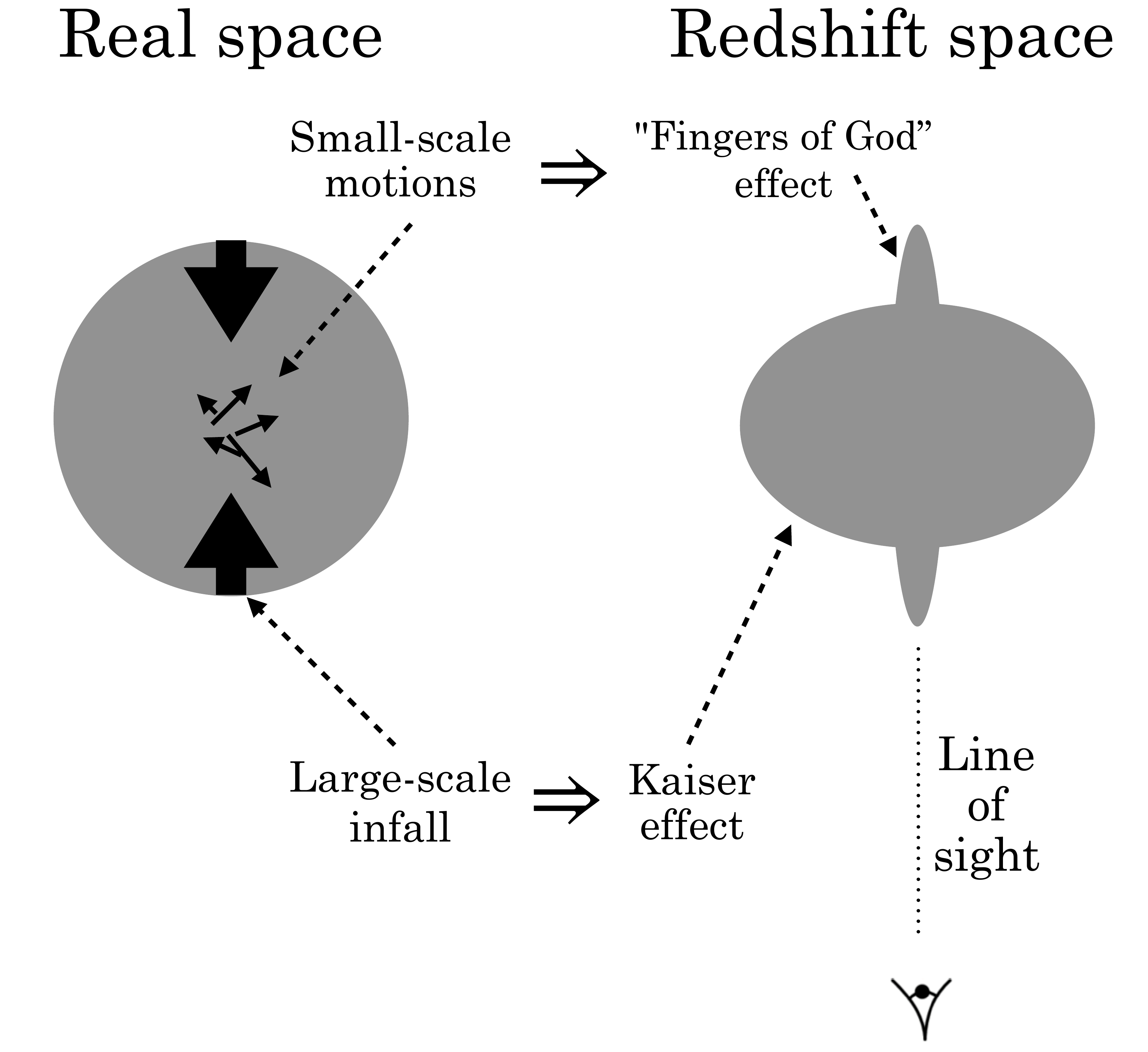}
\end{center}
\caption{Illustration of the redshift-space distortions. There are two
  principal effects: 1) On large scales, velocity flows into large
  overdensities, ``squishing'' the appearance of the object along the line of
  sight; this is the Kaiser effect; 2) On smaller scales, random motions
  introduce apparent elongation along the line of sight; this is the somewhat
  hyperbolically called ``fingers of god'' effect. Adopted from \cite{Huterer_book}. }
\label{fig:RSD}
\end{figure}

Probably the most prominent method for utilizing the velocities are the
redshift-space distortions (RSD). The RSD take place because redshift of
galaxies is affected by their gravitational infall into nearby large-scale
structures, as well as by the galaxies' own peculiar velocities. Because we
typically measure the galaxies' redshifts in order to get their radial
location, clustering measurements in redshift space are subject to RSD. There
are two RSD effects --- the so-called Kaiser effect, and fingers of god; see
Fig.~\ref{fig:RSD}. To lowest order, the redshift-space matter power spectrum $P(\bfk)^{(s)} $
is related to the isotropic matter power spectrum $P(k)$ via
(e.g.\ \cite{Song:2008qt})\footnote{A slightly more accurate formulation
  replaces the density power spectrum $P(k, z)\equiv P_{\delta\delta}$ on the
  right-hand side of Eq.~(\ref{eq:Kaiser_RSD}) with a combination of $P_{\delta\delta}$, the velocity power
  spectrum $P_{vv}$, and the cross-power $P_{\delta v}$, as \cite{delaTorre:2012dg}
  \begin{equation*}
  P(\bfk, z)^{(s)}  = \left [b^2P_{\delta\delta}+ 2bf\mu^2P_{\delta v} + f^2\mu^4P_{vv}\right ]
  F(k^2\sigma_v^2\mu^2).
  \end{equation*}
  }
\begin{equation}
  P(\bfk, z)^{(s)}  = \left [b+f\mu^2\right ]^2F(k^2\sigma_v^2\mu^2)P(k, z),
  \label{eq:Kaiser_RSD}
\end{equation}
where the superscript $\!^{(s)}$ indicates that we are referring to clustering
in redshift space.  Here $\mu$ is the cosine of the angle between the
line-of-sight direction and wavenumber $\bfk$, $b=b(k, z)$ is the galaxy bias,
$f=f(z)$ is the growth rate, and $P(k, z)$ is the isotropic part of the
matter power spectrum.  Finally, the function $F(k^2\sigma_v^2\mu^2)\simeq
1/(1+k^2\sigma_v^2\mu^2)$ models the suppression of the redshift-space power
spectrum at high $k$ in terms of the velocity dispersion $\sigma_v$, which
is a free parameter that generally depends on halo mass and redshift.

The RSD effects
can be readily measured in spectroscopic surveys where accurate galaxy
redshifts are available.  Because $P(\bfk)^{(s)}$ has terms that go as
$\mu^0$, $\mu^2$, and $\mu^4$ (in linear theory), we can respectively measure
the monopole, quadrupole, and hexadecapole of redshift-space galaxy
clustering.

An alternative to utilizing the RSD to probe growth is to measure the velocities
directly. Redshift to a galaxy is affected by its velocity, so accurate measurements of the
former can constrain the latter, and hence cosmic growth. Velocity of a galaxy
at a physical distance $\bfx$ is
\begin{equation}
\dot\bfx = \frac{d}{dt}(a\bfr)= \dot{a}\bfr + a\dot\bfr = H \bfx + \bfv_{\rm pec},
\end{equation}
where $\bfr$ is the comoving distance, $H$ is the Hubble parameter, and
$\bfv_{\rm pec}$ is the component of the peculiar velocity of the galaxy
parallel to the line of sight.  Specializing in $z\ll 1$ (and ignoring
2nd-order terms in redshift), we get the relation between the observed
redshift $z_{\rm obs}$ and true redshift $z$ as
\begin{equation}
cz_{\rm obs}\simeq cz + v_{\rm pec}.
\end{equation}
The effect of peculiar velocities is non-negligible only for very nearby galaxies;
consider that for a galaxy at $z\simeq 0.03$, $cz\simeq 10,000\kms$, while
the typical peculiar velocity is $v_{\rm pec}\simeq 200\kms$.

To determine the peculiar velocity, we need the observed redshift and 
distance. Galaxy redshift can be determined by taking a sufficiently
high-resolution spectrum of the galaxy, and is in principle straightforward.
Measuring the distance to a galaxy is far more challenging. To get the
distance, one may use an empirical fundamental-plane relation that relates an
object's distance to its surface brightness. Alternatively, one can use type
Ia supernovae, standard candles whose distances can be determined by measuring
their fluxes and adopting the fact that these objects have nearly equal
luminosities.  In surveys thus far, the fundamental-plane measurements are
available for $\sim$10,000 galaxies and give distances accurate to $20-30$\%
per object. Type Ia supernovae are more accurate ($\sim$10\% per object), but
available for fewer galaxies (typically $\lesssim 500$) in extant
peculiar-velocity surveys. Either method allows a noisy measurement of
individual galaxies' peculiar velocities; their correlations are sensitive to
growth as indicated in Eq.~(\ref{eq:P_vv}). Recent observational efforts,
combined with modern distance calibrations, have been greatly increasing the
number of galaxies and supernovae available for peculiar-velocity tests of cosmology
\cite{Howlett:2022len,Tully:2022rbj}.

For practical reasons having to do with patchy sky coverage and sparsity of
source galaxies, one often measures the (configuration-space)
\textit{correlation function} of galaxy velocities. The correlation function
$C_{vv}$ is proportional to the velocity power spectrum. Since the velocity power
spectrum is proportional to the quantity $\fs8\equiv f(z)\sigma_8(z)$
\begin{equation}
  P_{vv}(k, z)\propto
  f^2(z)P(k, z)= f^2(z) \sigma_8^2(z)P(k, z=0),
\end{equation}
so is the velocity correlation function
\begin{equation}
  C_{vv}\propto P_{vv} \propto  (f\sigma_8)^2.
  \label{eq:vv}
\end{equation}
This relation explains why the results from the analyses of cosmic velocities
are often framed as constraints on the quantity $f\sigma_8$, evaluated in
redshift bins. Note however that, while the strongest dependence of velocity
correlations is on the quantity $f\sigma_8$, most of the standard cosmological
parameters also enter, mainly via their impact on $P(k)$ that is not captured
by $\sigma_8$.  This complicates attempts to isolate the measurements on
$f\sigma_8$, as these other parameters (e.g.\ the physical baryon density
$\Omega_b h^2$ or the spectral index $n_s$) need to be given priors from other
data, or else be simultaneously constrained by the velocities.

Observing the peculiar velocities of nearby galaxies or type Ia supernovae
that they host is not the only way to probe the galaxy velocity field. One
useful probe in this regard is the \textit{kinetic SZ effect} (kSZ). The kSZ
is caused by galaxy clusters' peculiar velocities in the CMB rest frame. A
cluster with peculiar velocity $v_{\rm pec, \parallel}$ along the line of sight will
incur a density shift of the CMB temperature in the direction of the cluster
of
\begin{equation}
  \left (\frac{\delta T}{T}\right )_{\rm kSZ} \simeq -\tau_e\frac{v_{\rm pec, \parallel}}{c},
  \label{eq:kSZ}
\end{equation}
where $\tau_e$ is the optical depth for electron scattering in the
cluster. The kSZ modifies the blackbody spectrum of an object by the
temperature shift given in Eq.~(\ref{eq:kSZ}). For typical values
$\tau_e\simeq 0.01$ and $v_{\rm pec}\simeq 500\kms$, the kSZ fractional
temperature shift is on the order of the primordial temperature anisotropy
($\sim 10^{-5}$) and thus quite small. The kSZ effect was first detected about
a decade ago \citep{Hand:2012ui}, and its better mapping will allow us to
probe the velocity field of galaxy clusters, and hence the growth of cosmic
structure.

\subsection{Role of the CMB}

Being a snapshot of the universe at $z\simeq 1000$, the cosmic microwave
background (CMB) naively does not have much information about the growth of
structure. Specifically, we are most interested in the growth at relatively
low redshift, $z\lesssim 2$, where growth slowly transitions from being robust
in the dark-matter dominated regime, to being strongly suppressed as dark
energy starts to dominate\footnote{In the concordance $\Lambda$CDM model with 30\%
  matter and 70\% dark energy, the energy densities in these two components
  are equal at $z\simeq 0.33$.}, and we certainly do not expect that the CMB be informative
at such a low redshift. While technically correct, these expectations need to be
modulated with the fact that the CMB is nevertheless extremely important in
pinning down the initial amplitude of the power spectrum (that is, the
parameter $A_s$ in Eq.~(\ref{eq:Deltasq})). This constraint, along with the
\textit{present}-day clustering amplitude (measured as $\sigma_8$ or $S_8$),
allows a significant improvement in the precision of constraints on the growth
of structure.

CMB contributes in other ways. Notably, CMB {\it lensing} --- measured as
subtle displacements in the distribution of hot and cold spots on arcminute
scales --- is sensitive to the growth of structure. CMB lensing is described
by the power spectrum of the deflection signal of photons as they cross
large-scale structure traveling toward our detectors. Mathematically, the CMB
lensing power spectrum looks similar to the weak-lensing shear power spectrum
in Eq.~(\ref{eq:Pkappa_ij}); the main difference is that kernel of the shear
power spectrum peaks at $z\sim 0.5$, while that of CMB lensing peaks at $z\sim
3$. Hence, CMB lensing in principle helps probe the growth of structure at a
redshift higher than galaxy or weak-lensing surveys. The signal-to-noise of
CMB lensing is currently limited, but future measurements from surveys like
Simons Observatory and CMB-S4 may provide interesting constraints on
the growth of structure in their own right. Another promising direction is to
cross-correlate the CMB map with a galaxy map; the contribution of CMB lensing
should lead to a sufficiently high signal-to-noise to probe the growth of
structure
\citep{Hu:2002rm,Peacock:2018xlz,Wilson:2019brt,Krolewski:2019yrv,Garcia-Garcia:2021unp}.

\begin{figure}[!t]
\begin{center}
\includegraphics[scale=0.6]{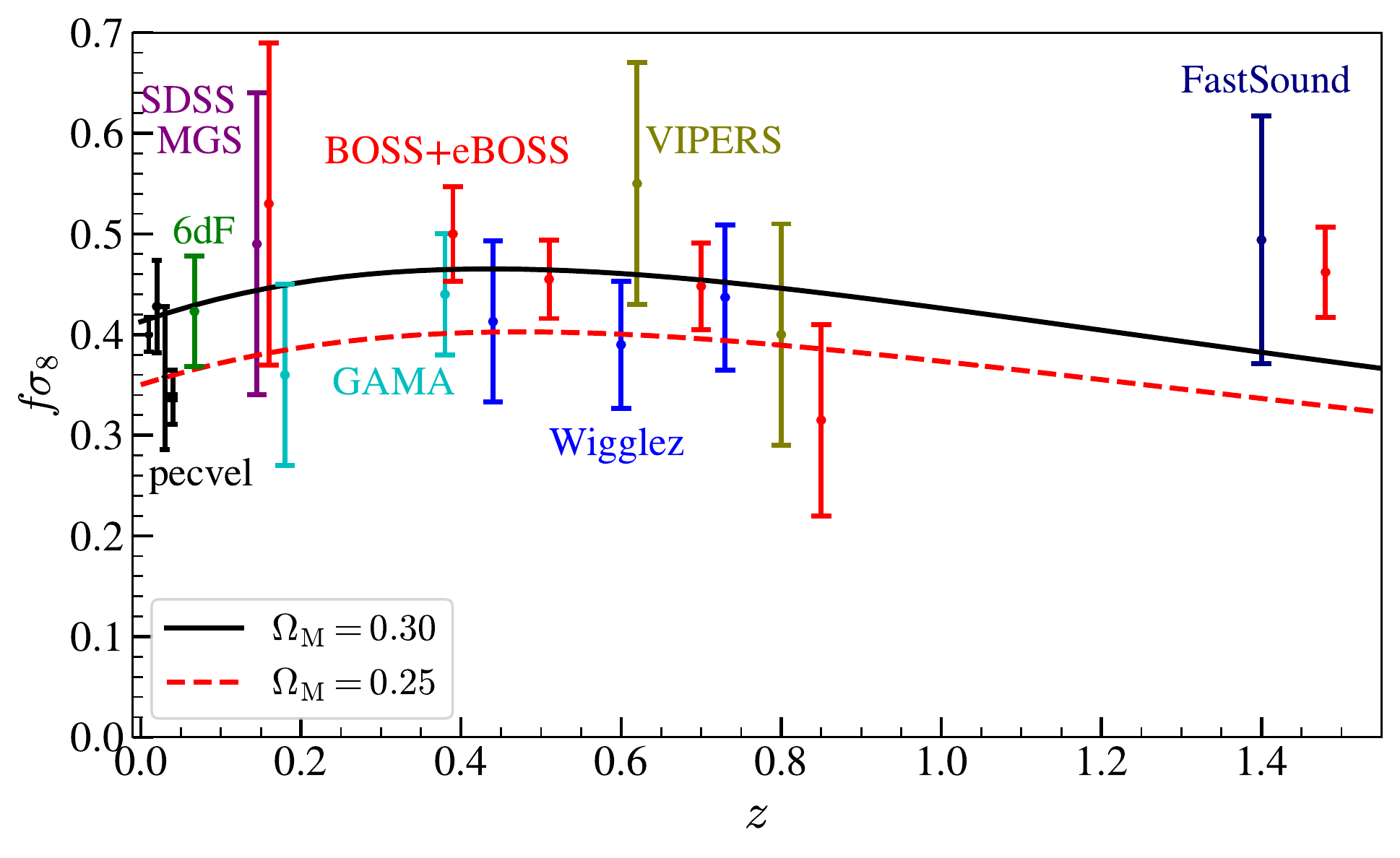}
\end{center}
\caption{Constraints on $\fs8\equiv f(z)\sigma_8(z)$. The curves show
  predictions from $\Lambda$CDM with two alternate values of the matter
  density --- the fiducial $\om=0.3$, and also $\om=0.25$. The
  errors show the various extant measurements, as follows. The black error
  bars denote constraints from peculiar velocities (or their combination with
  clustering) at $z\simeq 0$; these black points are mutually offset
  horizontally for clarity. From left to right, these peculiar-velocity
  constraints come from: type Ia supernovae from the 2M++ compilation
  \cite{Boruah:2019icj}, supernova distances from Supercal compilation and
  Tully-Fisher distances from 6dFGRS \cite{Huterer:2016uyq}, combination of
  galaxy velocities and galaxy clustering applied to 6dFGRS
  \cite{Turner:2022mla}, and combination of galaxy velocities from 6dFGRS and
  SDSS with velocity and density predictions from 2M++ galaxy survey
  \cite{Said:2020epb}. The other constraints, at higher redshift, come from
  redshift-space distortion modeling of galaxy samples; see text for details.
}
\label{fig:fs8}
\end{figure}

\subsection{Cosmological constraints on $\fs8$ and $S_8$}\label{sec:constraints}

Having reviewed the principal cosmological probes of growth, we now summarize
current constraints.  While the temporal evolution growth enters many of the
observable quantities in cosmology, isolating it is in general not
straightforward; we discuss such strategies in Section \ref{sec:consistency}
below.  Instead, the majority of the cosmological analyses assume that growth
is modeled according to the assumed cosmological model (e.g.\ $\Lambda$CDM and
its variants, or else modified gravity), and report the constraints on the
parameters of that model, say $\om$ and $\sigma_8$.  In order to present
results in a more ``model-independent'' way, these surveys report constraints
on derived cosmological parameters that are fairly directly related to what is
being measured, yet contain information  that in large part comes from
growth. We now discuss these constraints.

Perhaps the most direct constraint comes from peculiar-velocity and
redshift-space-distortion measurements, both of which are sensitive to the
quantity $\fs8$ at some effective redshift that depends on the distribution of
source galaxies (see Eqs.~(\ref{eq:P_vv}) and (\ref{eq:vv})). The quantity $\fs8$
depends on both growth and other cosmological parameters, principally via
how they enter the matter power spectrum. Despite some degeneracy with these
other parameters, the dependence on growth is very strong because both $f(z)$
and $\sigma_8(z)$ are directly related to the linear growth $D(z)$;
$\fs8\propto (d\ln D/d\ln a) D$.

Constraints on $\fs8$ from current measurements are shown in
Fig.~\ref{fig:fs8}. The black error bars denote constraints from peculiar
velocities (or their combination with clustering) at $z\simeq 0$; the points
are mutually offset horizontally for clarity. From left to right, these
peculiar-velocity constraints come from: type Ia supernovae from the 2M++
compilation \cite{Boruah:2019icj}, supernova distances from Supercal
compilation and Tully-Fisher distances from 6dFGRS \cite{Huterer:2016uyq},
combination of galaxy velocities and galaxy clustering applied to 6dFGRS
\cite{Turner:2022mla}, and combination of galaxy velocities from 6dFGRS and
SDSS with velocity and density predictions from 2M++ galaxy survey
\cite{Said:2020epb}. The other constraints, at higher redshift, are mainly
from redshift-space distortion modeling of galaxy samples. These RSD
constraints come from: 6dFGRS \citep{Beutler:2012px}, GAMA
\citep{Blake:2013nif}, WiggleZ \citep{Blake:2012pj}, VIPERS
\citep{Pezzotta:2016gbo}, SDSS main galaxy sample (MGS)
\citep{Howlett:2014opa}, FastSound on Subaru \citep{Okumura:2015lvp}, and
BOSS+eBOSS \citep{eBOSS:2020yzd}.  We see that the $\fs8$ constraints are in a good
agreement with the predictions of the currently favored $\Lambda$CDM
model. Particularly notable is the complementarity between the different
probes, as peculiar velocity surveys measure motions of galaxies at distances
$r\lesssim 100\hinvmpc$, and thus very low redshift ($z\simeq 0.02$), while
the RSD measurements probe growth to galaxies and quasars at a much higher
redshift, all the way up to $z\simeq 1.5$ for current surveys.

In contrast to peculiar-velocity and RSD measurements, constraints from
weak-lensing surveys and measurements of the broadband galaxy-clustering power
spectrum are not very sensitive to the growth rate $f$ or the combination $\fs8$, but rather on the
overall amplitude of matter fluctuations. This amplitude is larger if either
matter density $\om$ or the amplitude of mass fluctuations $\sigma_8$
increases. Therefore, the constraints typically look like a banana-shaped
region in the $\om-\sigma_8$ plane, indicating their mutual
anti-correlation. To decouple these two parameters, the constraints are often
reported on their combination $S_8\equiv \sigma_8 (\om/0.3)^{0.5}$ that
is very well constrained. Note that $\sigma_8$ and $S_8$ explicitly refer to
the value of these quantities at redshift zero, to which the weak-lensing and
galaxy-clustering are extrapolated by convention, in contrast to
peculiar-velocity and RSD constraints which are traditionally quoted as
$\fs8\equiv f(z)\sigma_8(z)$.

\begin{figure}[!t]
\begin{center}
\includegraphics[scale=0.5]{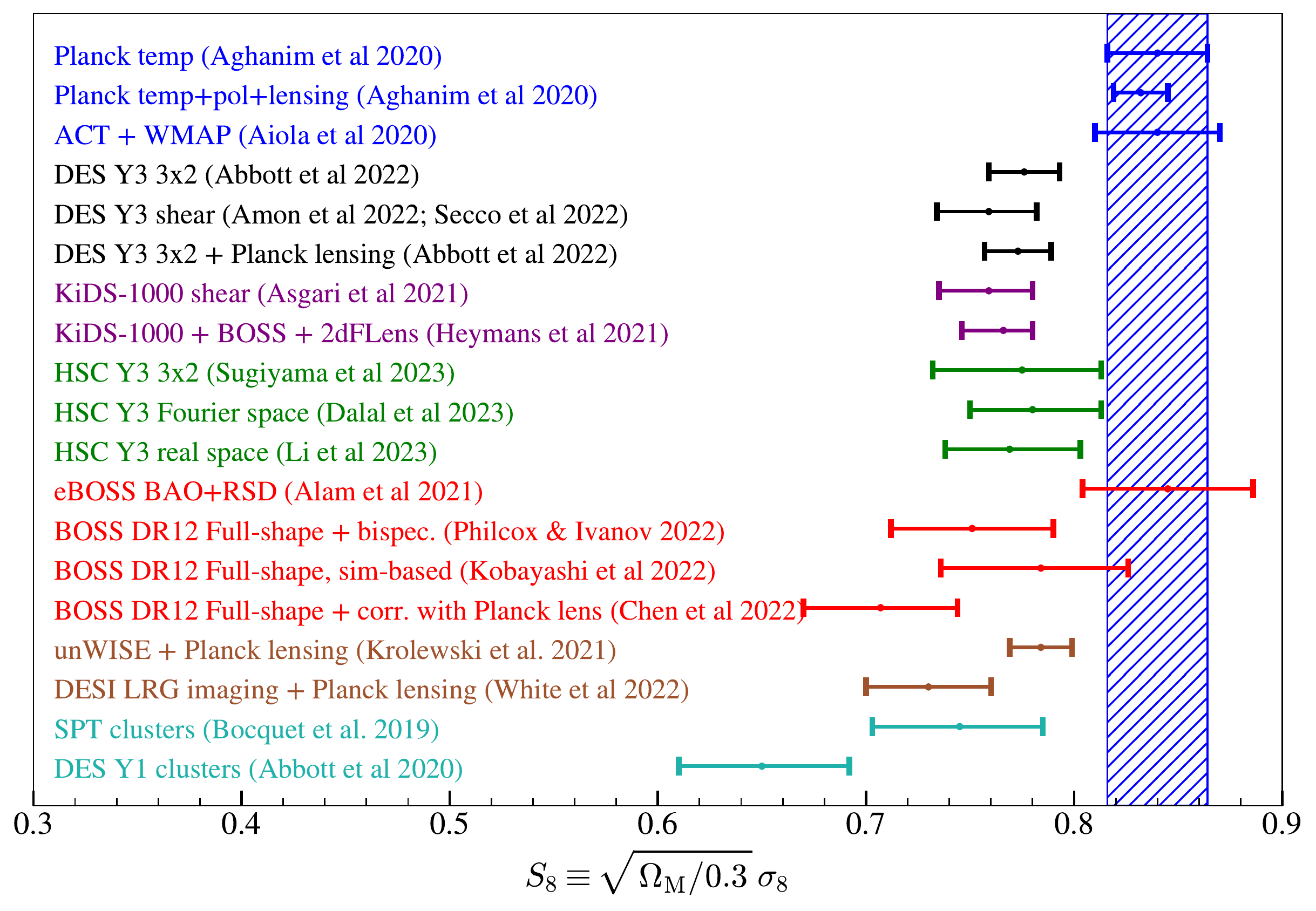}
\end{center}
\caption{Constraints on $S_8\equiv \sigma_8 \sqrt{\om/0.3}$ from various
  analyses. Going top to bottom, the first three measurements come from the
  CMB analysis: temperature-only analysis from Planck
  (\cite{Planck:2018vyg}, also shown as a vertical shaded region), temperature
  + polarization + lensing analyses from Planck \cite{Planck:2018vyg}, and
  temperature + polarization information from ACT combined with large-scale
  temperature from WMAP \cite{ACT:2020gnv}. All of the other constraints
  involve some combination of large-scale structure probes, see text for a
  detailed description. All
  constraints assume the fiducial $\Lambda$CDM model.
  The error bars show
    68\% credible intervals, which are asymmetric around the mean in some
    cases.}
\label{fig:S8}
\end{figure}

Figure \ref{fig:S8} shows the constraints on $S_8$. Going top to bottom, the
first three measurements come from CMB analyses: temperature-only analysis
from Planck (\cite{Planck:2018vyg}; also shown as a vertical shaded region),
temperature+polarization+lensing analysis from Planck \cite{Planck:2018vyg},
and temperature+polarization information from ACT combined with large-scale
temperature from WMAP \cite{ACT:2020gnv}. Going further down, we show the
constraints of select large-scale structure combinations as follows.  The
black error bars show constraints from the fiducial 3x2 analysis from DES Y3
\cite{DES:2022ccp}, DES Y3 with cosmic shear alone
\cite{DES:2021bvc,DES:2021vln}, and DES 3x2 with the addition of Planck
lensing information \cite{DES:2022urg}. The purple error bars show constraints
from KiDS-1000 \cite{KiDS:2020suj} and KiDS-1000 combined with clustering from
BOSS and galaxy-galaxy lensing in the overlap region of KiDS, BOSS and 2dFLens
\cite{Heymans:2020gsg}. The green error bars show the constraints from the
Hyper Suprime-Cam (HSC) year-3 data analysis: their 3x2 analysis result
\cite{Sugiyama:2023fzm}, and their Fourier-space \cite{Li:2023tui} and
real-space \cite{Dalal:2023olq} shear-only constraints. The red error bars show
constraints from the combined BAO and RSD information from BOSS+eBOSS
\cite{eBOSS:2020yzd}, from the emulator-based full-shape analysis from BOSS DR
12 \cite{Kobayashi:2021oud}, full-shape BOSS DR12 analysis combined with the
BOSS DR12 bispectrum monopole \cite{Philcox:2021kcw}, and full-shape BOSS DR12
along with its cross-correlations with Planck lensing \cite{Chen:2022jzq}. The
brown error bars show combination of unWISE galaxy clustering and Planck
lensing \cite{Krolewski:2021yqy}, and clustering of luminous red galaxies from
DESI imaging survey combined with Planck lensing \cite{White:2021yvw}. The
turquoise error bars show constraints from the abundance of galaxy clusters as
measured by South Pole Telescope \cite{SPT:2018njh} and DES Year-1 data
\cite{DES:2020ahh}.  All constraints assume the fiducial $\Lambda$CDM model.

The most apparent trend in Fig.~\ref{fig:S8} is the so-called\footnote{Also
  referred to as the ``$\sigma_8$ tension'', as a similar trend is seen in
  constraints on $\sigma_8$.} ``S8 tension'', which reflects the fact that CMB
measurements show a higher amplitude of mass fluctuations than lensing
surveys. Planck's temperature, polarization, and lensing information combined
indicate $S_8=0.832\pm 0.013$, and the combination of ACT and WMAP give a
consistent result ($S_8=0.840\pm 0.040$). Lensing surveys, on the other hand,
typically show a lower value of this parameter, exemplified by DES Y3
constraint $S_8=0.775\pm 0.017$ and a very similar constraint from KiDS
($S_8=0.759^{+0.024}_{-0.021}$). Similar trends are seen with cluster
abundance constraints. Future data will sharply improve the constraints on 
growth.

\begin{figure}[!t]
\begin{center}
\includegraphics[scale=0.70]{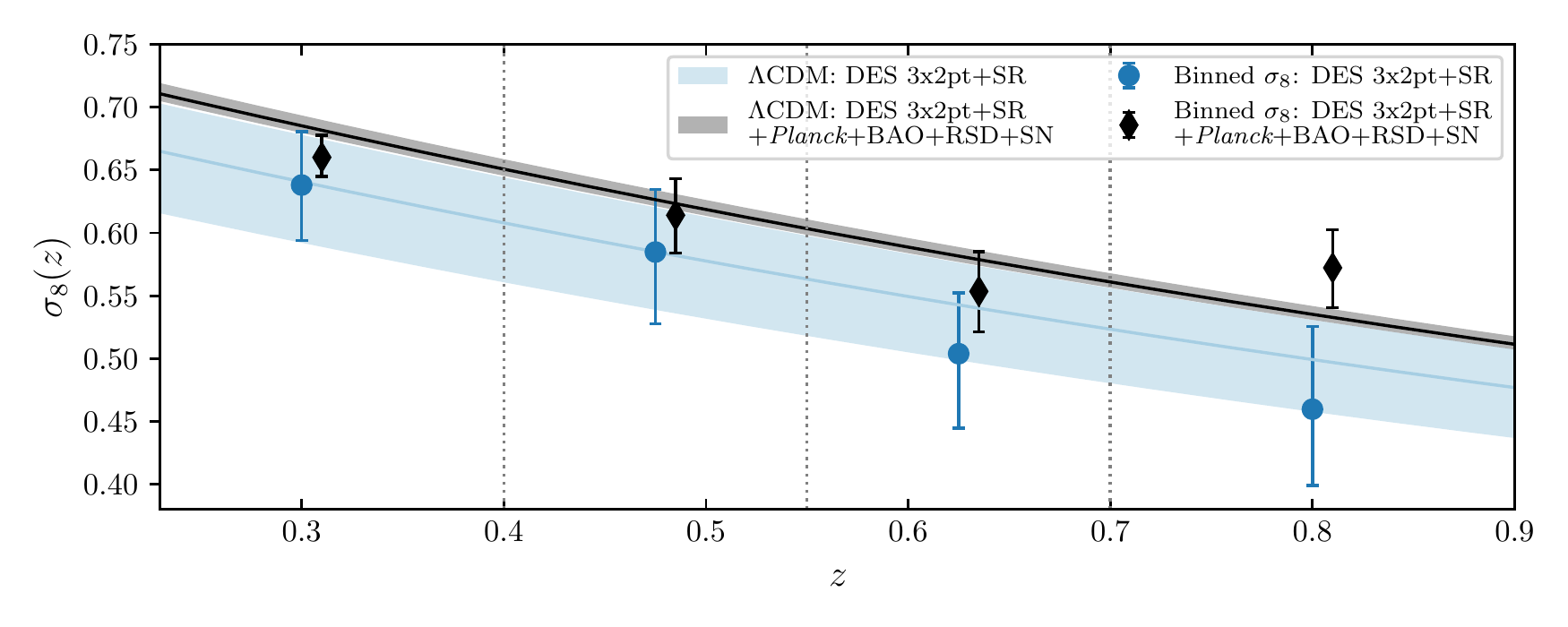}
\end{center}
\caption{Constraints on the time-dependent amplitude of mass fluctuation
  $\sigma_8(z)$ and comparison with theory, adopted from DES Y3 extensions
  paper \cite{DES:2022ccp}. The blue points show the constraints from DES
  3$\times$2 analysis alone in each of the four lens redshift bins. The black
  points show constraints from the combination of DES 3$\times$2 and external
  data (Planck, BAO, RSD, SN), along with the shear ratio statistic from DES
  (labeled SR).  Lines and shaded bands show the means and 68\% credible
  intervals inferred from $\Lambda$CDM posteriors corresponding respectively to
  DES alone (blue), and DES plus external data (black). }
\label{fig:sigma8_z}
\end{figure}

Constraints on $\sigma_8$ or $S_8$ are typically projected to, and reported at,
redshift zero. One can do better however, and constrain the temporal evolution
of the (linear) amplitude of mass fluctuations, with the understanding that
these amplitude parameters are directly proportional to the linear growth function
$D(z)$. Such constraints are shown in Figure \ref{fig:sigma8_z}, adopted from
DES Y3 extended-model paper \cite{DES:2022ccp}. The blue points show the
constraints from DES 3$\times$2 analysis alone in each of the four lens
redshift bins. The black points show constraints from the combination of DES
and external data: CMB, baryon acoustic oscillations (BAO), redshift-space
distortions (RSD), and type Ia supernovae (SN).  Lines and shaded bands show the
mean and 68\% credible interval inferred from $\Lambda$CDM posteriors
corresponding respectively to DES alone, and DES plus external data. Clearly,
$\sigma_8(z)$ measurements are already accurate, especially when various
datasets are combined, and sharply test (and, thus far, are in agreement with)
the standard $\Lambda$CDM cosmological model.

\section{Consistency tests with growth}\label{sec:consistency}

As mentioned in the Introduction, growth of structure owes its outsize
importance chiefly to the fact that it is a powerful discriminator between
models of dark energy, and that it is sensitive to the presence of modified
gravity (e.g.\ \cite{Lue:2003ky,Amendola:2007rr}). In Section \ref{sec:obs} we discussed how individual probes are
sensitive to growth, and in particular (in Sec.~\ref{sec:constraints}) how the
contraints derived from growth are reported in the measurements of $\fs8(z)$
and $S_8$.  We now discuss more ambitious and direct uses of the information
stored in the growth of structure, specifically how to utilize it to test
classes of cosmological models.

\subsection{Constraints on parameterized growth}\label{sec:gamma_growth}

While it is possible to isolate the constraints coming specifically from the
growth of cosmic structure (as we discuss below in Sec.~\ref{sec:grow-geom}),
this is typically not straightforward to implement. Moreover, such relatively
general extractions of the growth information necessarily impose additional parameters that
weaken the cosmological constraints.  This is where a \textit{simple}
parameterization of growth becomes extremely useful. A successful such
parameterization would have the ability to describe growth in $\Lambda$CDM and
wCDM (where the equation of state of dark energy $w$ is a free but constant
parameter), as well as model departures expected in modified-gravity
scenarios.

By far the most impactful parameterization of this
kind is provided by ``growth index'' parameter\footnote{Not to be
  confused with shear which we discussed in Sec.~\ref{sec:shear}.} $\gamma$. This
description of growth introduces a single, constant parameter defined in a
phenomenological fit to the growth rate \cite{Linder:2005in}
\begin{equation}
  f(a)\equiv \om(a)^\gamma.
  \label{eq:index}
\end{equation}
In other words, the linear growth factor is approximated by
$D(a)= e^{\int_0^a d\ln a'\,[\om(a')^\gamma]}$.

It has been known for a long time that the formula in Eq.~(\ref{eq:index})
fits the linear growth very well for $\gamma\simeq 0.55$
(e.g.\ \cite{Peebles:1994xt,Wang:1998gt}). What is new here is that promoting
$\gamma$ to a free parameter enables describing growth in models far beyond the
standard $\Lambda$CDM.  Specifically, it has been shown that dynamical dark
energy models where the equation of state ratio is parametrized as
$w(a)=w_0+w_a(1-a)$ are well fit by the growth index as long as it takes the
value \cite{Linder:2005in}
\begin{equation}
  \gamma = 0.55 + 0.02 [1+w(z=1)],
  \label{eq:gamma_w0wa}
\end{equation}
where $w(z=1)$ is the dark-energy equation of state evaluated at redshift one in this
class of models. The form in Eq.~(\ref{eq:gamma_w0wa}) fits the exact linear
rate to better than 0.3\% when $w_0+w_a<-0.1$ \cite{Huterer:2006mva}. A broad
range of modified-gravity models, including time-varying gravity, DGP braneworld
gravity \cite{Dvali:2000hr}, and some scalar-tensor gravity, are fit accurately with
the growth index \cite{Linder:2007hg}.


There are other variants of parametrized growth. A more direct, and even less
model-dependent, approach is to model $D(z)$ as a free function in redshift
and interpolate it with principal components \cite{Hu:2002rm},
splines \cite{Garcia-Garcia:2021unp}, or else with piecewise-constant values as
done for $\sigma_8(z)$ in the DES Y3 extensions paper (\cite{DES:2022ccp};
see our Fig.~\ref{fig:sigma8_z}).  As the measurements of the growth of
structure become more precise with upcoming surveys, such ambitious approaches
will begin to return very interesting constraints and consistency-test
results.

\subsection{Comparing growth with geometry}\label{sec:grow-geom}

Comparing measurements of geometric quantities to those describing the
growth of structure is a particularly promising stress-test of the
cosmological model
\cite{Zhang:2003ii,Bernstein:2003es,Ishak:2005zs,Knox:2005rg,Bertschinger:2006aw,Huterer:2006mva}.
For example, given constraints on the initial conditions (power spectrum shape
and amplitude), very precise distance measurements from e.g.\ type Ia
supernovae (SNe Ia) and baryon acoustic oscillations (BAO) predict the
convergence power spectrum measured by weak lensing probes. Here, the weak
lensing signal depends on the late-time growth of structure, which in turn is
precisely determined by distance measurements.

A straightforward way to compare geometry and growth, first proposed in the
modern form and applied to early data by \cite{Wang:2007fsa}, and then further
developed by \cite{Ruiz:2014hma,
  Bernal:2015zom,DES:2020iqt,Ruiz-Zapatero:2021rzl,Andrade:2021njl}, is to add
additional cosmological parameters. Specifically, in the flat $\Lambda$CDM
cosmological model one can take the matter density relative to the critical
--- which is normally encoded in one parameter, $\om$ --- and duplicate
it into two parameters: $\omgeom$ and $\omgrow$. One then modifies the theory
code as follows: every geometrical term (for example, a distance formula) will
depend on $\omgeom$, and every growth term (for example, in the
growth-of-structure differential equation) will be fed $\omgrow$. With such an
implementation, the tight relations linking geometry to growth in the standard
cosmological model will be explicitly decoupled.  In a flat cosmological
model, the energy density of dark energy is given by
$\Omega_\Lambda=1-\om$, so such a split of $\om$ is also
automatically a split of $\Omega_\Lambda$.
One can extend such a geometry-growth split to models with more complicated
dark-energy sectors. For example, in the analogous scenario of the flat wCDM
model, the equation-of-state of dark energy, $w$, is also described by two
parameters, $\wgeom$ and $\wgrow$. Therefore, a cosmological analysis with
geometry-growth split is specified by
\begin{flalign}
  \qquad \bullet\ \mbox{split\,\,} \Lambda{\rm CDM:} &\quad \{\omgeom, \omgrow, \{p_i\}\}&  \label{paramsom} \\[0.2cm]
  \qquad \bullet\ \mbox{split\,\,} w{\rm CDM:} &\quad \{\omgeom, \omgrow, \wgeom, \wgrow, \{p_i\}\} \label{paramsomw}
\end{flalign}
where $\{p_i\}$ are other, standard (and unsplit) cosmological parameters.
Note that, while the implementation of the geometry-growth split may be
  ambiguous (i.e.\ it might be unclear whether a given term in a theory
  equation is ``geometry'' or ``growth''), the geometry-growth test is always
  valid. This is because \textit{any} mismatch between geometry and growth
  terms --- however they are implemented --- is disallowed in the standard
  (unsplit) cosmological model\footnote{Of course, it is a good idea to implement
  the separation between geometry and growth terms  that is physically
  sensible.}.

\begin{figure}[t]
\begin{center}
\includegraphics[scale=0.45]{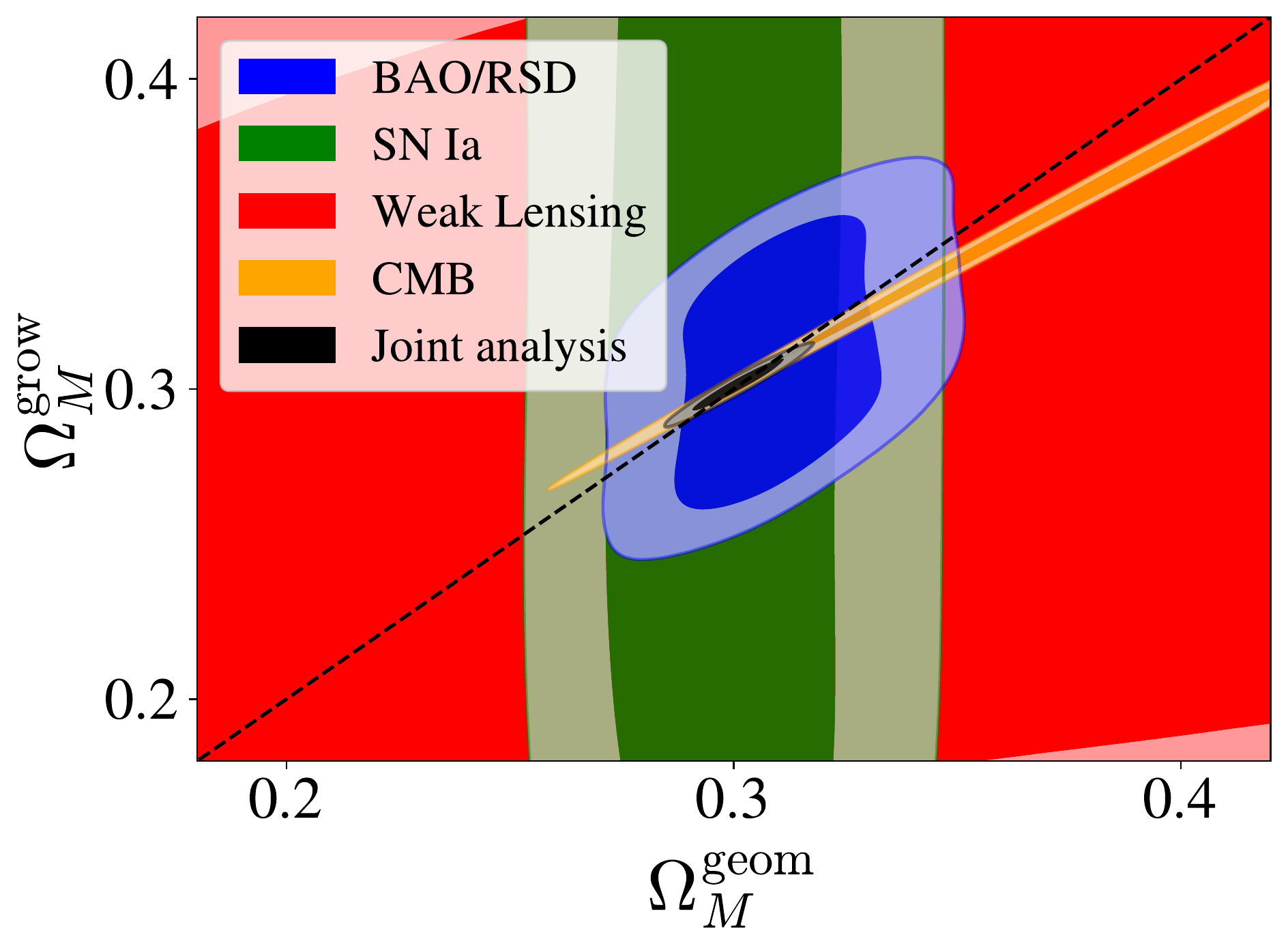}
\end{center}
\caption{Geometry-growth constraints from individual probes as well as the
  joint analysis. Here we extend the wCDM model by assuming two parameters for
  the matter density relative to critical, $\omgeom$, which governs the
  gemetrical quantities, and $\omgrow$, which controls the growth of
  structure; we also split the equation of state into $\wgeom$ and $\wgrow$,
  and marginalize over both of these parameters (and all others, unsplit
  parameters) in this Figure.  The dashed black line shows the consistency relation that is
  satisfied in the standard cosmological model, $\omgeom = \omgrow$. Note the
  impressive degeneracy breaking when the cosmological probes are
  combined. Adopted from Ref.~\cite{Andrade:2021njl}.}
\label{fig:split}
\end{figure}

Constraints on the geometry and growth parameters in the split wCDM model are
shown in Fig.~\ref{fig:split}, adopted from \cite{Andrade:2021njl}.  Here we
show the constraints on $\omgeom$ and $\omgrow$, marginalized over all other
parameters, including $\wgeom$ and $\wgrow$. Note that SN Ia are manifestly
sensitive to geometry only, and not to growth. The constraints from individual
probes are necessarily weak, as the split parameter space contains severe
parameter degeneracies. When the probes
are combined, however, parameter degeneracies are broken and the constraints
are strong.

Current constraints on geometry and growth indicate intriguing departures from
the standard cosmological model (which predicts $\omgeom =\omgrow$ and $\wgeom
=\wgrow$) with some combinations of datasets
\cite{Ruiz:2014hma,Bernal:2015zom,Andrade:2021njl}. These results will be well
worth following up with forthcoming data.

\subsection{Falsifying classes of models}

Another very effective way to test the consistency of any given cosmological
model is to:
\begin{itemize}
  \item Start with constraints on the model that are provided by certain datasets
    or probes;
  \item Compute what those parameter constraints imply for the larger space
    of observable quantities, for example cosmological functions $D(z)$, or
    $\fs8(z)$ at an arbitrary redshift, or else derived parameters such as $S_8$; and  then
\item Test \textit{those}   predictions using new data or probes.
\end{itemize}
Such an approach has implicitly been discussed for a long time
(e.g.\ \cite{Huterer:2006mva}), but was first clearly spelled out by Mortonson
et al.\ \cite{Mortonson:2008qy} and applied to current data by
Refs.~\cite{Mortonson:2009hk,Vanderveld:2012ec,Miranda:2017mnw,Raveri:2019mxg}.

Falsification procedure as described above has a few clear advantages. First,
the simple structure of smooth dark-energy models implies that, given current
constraints, these models lead to very precise predictions for the temporal
evolution of distance and growth at \textit{all} redshifts, even well beyond
the range probed by data used to impose the predictions. Second, those
predictions are quite accurate not only for $\Lambda$CDM and wCDM, but 
also for models with a much richer dark-energy sector
\cite{Mortonson:2008qy}. This makes the predictions good targets for falsifying
whole classes of models, particularly with data that probe new ranges of
redshift (or new spatial scales) than those used to impose the
predictions. Third, the ability to separately test growth and geometry comes
about naturally in these tests, as predictions for both quantities as a
function of redshift can be made straightforwardly, essentially by running the
standard theoretical calculation constrained to the range of cosmological
parameters allowed by current data.

We have already shown one example of predictions in this review; this is the
Planck measurement of $S_8$, shown as the vertical band in
Fig.~\ref{fig:S8}. Note that the spirit of model predictions is at work here:
even though $S_8$ is not a core parameter that CMB constrains, its value can
be easily predicted by the CMB anisotropy measurements in a given
cosmological model (here, $\Lambda$CDM).  This prediction serves as a check
against more direct $S_8$ measurements from galaxy clustering and weak lensing.

There are other ways to falsify models that are not quite as direct as the 
procedure described above. One such test is to compare the two gravitational
potentials $\Psi$ and $\Phi$. Starting from the scalar-perturbed
Friedmann-Robertson-Walker line element in the conformal Newtonian gauge
\begin{equation}
ds^2 = a^2(\eta) \left[ (1 + 2\Psi) d\eta^2 - (1 - 2\Phi) \delta_{ij} dx_i dx_j \right],
\end{equation}
the behavior of many cosmologically interesting modified-gravity theories can
be fit with a free function of time and scale multiplying the Poisson
equation, and another given by the ratio of $\Phi$ and $\Psi$. Specifically,
one introduces two new functions of time\footnote{We do not allow scale
dependence of $\mu$ and $\Sigma$ in the following discussion, although such
scale dependence may be observable with future data; see
e.g.~\cite{Hojjati:2013xqa}.}  $\mu(a)$ and $\Sigma(a)$, defined as
\begin{equation}
  \begin{aligned}
k^2 \Psi &=  -  4 \pi G a^2 (1+\mu(a)) \rho \delta  \\[0.2cm]
k^2 (\Psi + \Phi) &=  - 8 \pi G a^2 (1+\Sigma(a))  \rho \delta \,,
\end{aligned}
\end{equation}
where $\delta$ is the comoving-gauge density perturbation. Here, $\Sigma$
parametrizes the change in the lensing effect on massless particles for a
given matter field, while $\mu$ describes the change in the matter overdensity
itself. Such a parameterization approximately describes a more complicated set
of equations \cite{Baker:2012zs, Creminelli:2008wc, Baker:2011jy,
  Battye:2012eu, Gleyzes:2013ooa, Gleyzes:2014qga, Bloomfield:2012ff,
  Amendola:2013qna, Hojjati:2013xqa}, but has been numerically verified on
scales of cosmological interest \cite{Noller:2013wca, Schmidt:2009sg,
  Zhao:2010qy, Barreira:2014kra, Li:2013tda}. Typically, one either assumes
that $\mu$ and $\Sigma$ are constant, or else that they scale with cosmic time
under some parameterization
\cite{Caldwell:2007cw,DES:2018ufa}. Finally, there exist
other, closely related, parameterizations of the gravitational potentials,
including the $E_G$ parameterization \cite{Zhang:2007nk} and parameterized
post-Friedmannian framework \cite{Hu:2007pj}.

Because $\Sigma$ determines the predictions for lensing, weak lensing
measurements are primarily sensitive to this parameter but also have some
smaller degree of sensitivity to $\mu$ via their tracing of the matter
field. Conversely, galaxy clustering measurements depend only on $\mu$ and are
insensitive to $\Sigma$ \cite{DES:2018ufa}. Constraints on $(\mu, \Sigma)$, or
similar quantities but with different names, are already very good
\cite{Hojjati:2015ojt,Salvatelli:2016mgy,Mueller:2016kpu,DES:2018ufa}.  More ambitious
analyses promote $\mu$ and $\Sigma$ to few-parameter functions of redshift
\cite{Daniel:2012kn,Simpson:2012ra, Ade:2015rim, Planck:2018vyg,DES:2018ufa,DES:2022ccp}.
Comparison of $\mu(a)$ and $\Sigma(a)$ to their fiducial values of zero thus
constitutes a test of modified gravity.

\section{Conclusions}\label{sec:concl}

We have described how the growth of structure determines key aspects of
observable quantities in cosmology, such as the spatial correlation of
galaxies, the coherence of their peculiar velocities, the amplitude and
spatial dependence of cosmic shear, the abundance of galaxy clusters, and
lensing of the cosmic microwave background (CMB) photons. Observations
of these quantities can provide important constraints on growth, which in turn
helps constrain cosmological models, and specifically distinguishes between
dark energy and modified gravity.

Notably, current constraints on the amplitude of mass fluctuations $S_8$ and
the function $\fs8$ come largely from growth. They can be used to look for
consistency of data with a cosmological model, as the $S_8$ and $\fs8(z)$
constraints are model-dependent. These constraints can also be used to look
for internal consistency between different probes or datasets, as well as
comparison between early- and late-universe constraints. Such comparisons have
been very fruitful, and have led to the currently much-discussed $S_8$
tension, where the  CMB  experiments appear to give a
higher value of this quantity than cosmic-shear measurements.

We highlighted several ways in which growth can be used to probe the
cosmological model that go beyond simply measuring and reporting $\sigma_8,
S_8$ or $\fs8(z)$. Explicit parameterization of the growth sector is one such
approach. For example, departures from growth predicted assuming general
relativity can be enabled by freeing the growth index $\gamma$ from its
general-relativistic value of $\simeq 0.55$, or by parameterizing the
deviations of gravitational potentials $\Psi$ and $\Phi$ from their fiducial values. One can
also encode additional freedom in the linear growth function $D(a)$ as a
function of scale or time. Alternatively, one can explicitly separate
information contained in the growth of structure from that in geometrical
terms by introducing separate geometry and growth parameters in equations that
govern cosmological observables. Because relations between the geometrical and
growth quantities that are obeyed in LCDM and wCDM are likely to be broken in
modified-gravity explanations of the accelerating universe, such
``geometry-growth split'' tests are potentially very powerful. Finally, we
discussed how simply making predictions given by measurements by one set of
probes (e.g.\ the CMB) for what another set of measurements (e.g.\ cosmic
shear or some other, new probe, or an existing probe in a new range of scales
or redshifts) is expected to measure, is also a very useful way to employ
growth to stress-test the cosmological model.

Most of the major upcoming experiments and telescopes with the role to probe
the accelerating universe rely on better measurements of growth, particularly
at higher redshifts ($z\simeq 2$-$5$) than currently possible, in order to
better understand dark energy and dark matter (see \cite{Annis:2022xgg} for a
recent review). Therefore, we expect that the measurements and
interpretations of the growth of structure will become even more central to
cosmology in the years to come.

\section*{Acknowledgements}

I would like to thank Eric Linder and GongBo Zhao for helpful comments on the
manuscript. Over the couple of years over which this manuscript evolved, my
work has been supported by NASA under contract 19-ATP19-0058, DOE under
Contract No.\ DE-FG02-95ER40899, NSF under contract AST-1812961, and Leinweber
Center for Theoretical Physics at the University of Michigan. I would also
like to thank Aspen Center for Physics and Max Planck Institute for
Astrophysics for their hospitality.


\bibliography{refs}

\begin{thebibliography}{100}
\providecommand{\url}[1]{{#1}}
\providecommand{\urlprefix}{URL }
\providecommand{\doi}[1]{\url{https://doi.org/#1}}
\bibcommenthead

\bibitem{Weinberg:2013agg}
D.H. Weinberg, M.J. Mortonson, D.J. Eisenstein, C.~Hirata, A.G. Riess, E.~Rozo,
  {Observational Probes of Cosmic Acceleration}.
\newblock Phys. Rept. \textbf{530}, 87--255 (2013).
\newblock \doi{10.1016/j.physrep.2013.05.001}.
\newblock {\href{https://arxiv.org/abs/1201.2434}{{arXiv:1201.2434}}}
  {[astro-ph.CO]}

\bibitem{Huterer:2013xky}
D.~Huterer, et~al., {Growth of Cosmic Structure: Probing Dark Energy Beyond
  Expansion}.
\newblock Astropart. Phys. \textbf{63}, 23--41 (2015).
\newblock \doi{10.1016/j.astropartphys.2014.07.004}.
\newblock {\href{https://arxiv.org/abs/1309.5385}{{arXiv:1309.5385}}}
  {[astro-ph.CO]}

\bibitem{2010gfe..book.....M}
H.~{Mo}, F.C. {van den Bosch}, S.~{White}, \emph{{Galaxy Formation and
  Evolution}} (2010)

\bibitem{Huterer:2023mmv}
D.~Huterer, \emph{{A Course in Cosmology}} (Cambridge University Press, 2023).
\newblock \doi{10.1017/9781009070232}

\bibitem{Zhao:2008bn}
G.B. Zhao, L.~Pogosian, A.~Silvestri, J.~Zylberberg, {Searching for modified
  growth patterns with tomographic surveys}.
\newblock Phys. Rev. D \textbf{79}, 083,513 (2009).
\newblock \doi{10.1103/PhysRevD.79.083513}.
\newblock {\href{https://arxiv.org/abs/0809.3791}{{arXiv:0809.3791}}}
  {[astro-ph]}

\bibitem{Zhao:2009fn}
G.B. Zhao, L.~Pogosian, A.~Silvestri, J.~Zylberberg, {Cosmological Tests of
  General Relativity with Future Tomographic Surveys}.
\newblock Phys. Rev. Lett. \textbf{103}, 241,301 (2009).
\newblock \doi{10.1103/PhysRevLett.103.241301}.
\newblock {\href{https://arxiv.org/abs/0905.1326}{{arXiv:0905.1326}}}
  {[astro-ph.CO]}

\bibitem{Daniel:2010yt}
S.F. Daniel, E.V. Linder, {Confronting General Relativity with Further
  Cosmological Data}.
\newblock Phys. Rev. D \textbf{82}, 103,523 (2010).
\newblock \doi{10.1103/PhysRevD.82.103523}.
\newblock {\href{https://arxiv.org/abs/1008.0397}{{arXiv:1008.0397}}}
  {[astro-ph.CO]}

\bibitem{Zhao:2010dz}
G.B. Zhao, T.~Giannantonio, L.~Pogosian, A.~Silvestri, D.J. Bacon, K.~Koyama,
  R.C. Nichol, Y.S. Song, {Probing modifications of General Relativity using
  current cosmological observations}.
\newblock Phys. Rev. D \textbf{81}, 103,510 (2010).
\newblock \doi{10.1103/PhysRevD.81.103510}.
\newblock {\href{https://arxiv.org/abs/1003.0001}{{arXiv:1003.0001}}}
  {[astro-ph.CO]}

\bibitem{Song:2010fg}
Y.S. Song, G.B. Zhao, D.~Bacon, K.~Koyama, R.C. Nichol, L.~Pogosian,
  {Complementarity of Weak Lensing and Peculiar Velocity Measurements in
  Testing General Relativity}.
\newblock Phys. Rev. D \textbf{84}, 083,523 (2011).
\newblock \doi{10.1103/PhysRevD.84.083523}.
\newblock {\href{https://arxiv.org/abs/1011.2106}{{arXiv:1011.2106}}}
  {[astro-ph.CO]}

\bibitem{Silvestri:2013ne}
A.~Silvestri, L.~Pogosian, R.V. Buniy, {Practical approach to cosmological
  perturbations in modified gravity}.
\newblock Phys. Rev. D \textbf{87}(10), 104,015 (2013).
\newblock \doi{10.1103/PhysRevD.87.104015}.
\newblock {\href{https://arxiv.org/abs/1302.1193}{{arXiv:1302.1193}}}
  {[astro-ph.CO]}

\bibitem{Yoo:2009au}
J.~Yoo, A.L. Fitzpatrick, M.~Zaldarriaga, {A New Perspective on Galaxy
  Clustering as a Cosmological Probe: General Relativistic Effects}.
\newblock Phys. Rev. D \textbf{80}, 083,514 (2009).
\newblock \doi{10.1103/PhysRevD.80.083514}.
\newblock {\href{https://arxiv.org/abs/0907.0707}{{arXiv:0907.0707}}}
  {[astro-ph.CO]}

\bibitem{Yoo:2010ni}
J.~Yoo, {General Relativistic Description of the Observed Galaxy Power
  Spectrum: Do We Understand What We Measure?}
\newblock Phys. Rev. D \textbf{82}, 083,508 (2010).
\newblock \doi{10.1103/PhysRevD.82.083508}.
\newblock {\href{https://arxiv.org/abs/1009.3021}{{arXiv:1009.3021}}}
  {[astro-ph.CO]}

\bibitem{Challinor:2011bk}
A.~Challinor, A.~Lewis, {The linear power spectrum of observed source number
  counts}.
\newblock Phys. Rev. D \textbf{84}, 043,516 (2011).
\newblock \doi{10.1103/PhysRevD.84.043516}.
\newblock {\href{https://arxiv.org/abs/1105.5292}{{arXiv:1105.5292}}}
  {[astro-ph.CO]}

\bibitem{Jeong:2011as}
D.~Jeong, F.~Schmidt, C.M. Hirata, {Large-scale clustering of galaxies in
  general relativity}.
\newblock Phys. Rev. D \textbf{85}, 023,504 (2012).
\newblock \doi{10.1103/PhysRevD.85.023504}.
\newblock {\href{https://arxiv.org/abs/1107.5427}{{arXiv:1107.5427}}}
  {[astro-ph.CO]}

\bibitem{Bonvin:2014owa}
C.~Bonvin, {Isolating relativistic effects in large-scale structure}.
\newblock Class. Quant. Grav. \textbf{31}(23), 234,002 (2014).
\newblock \doi{10.1088/0264-9381/31/23/234002}.
\newblock {\href{https://arxiv.org/abs/1409.2224}{{arXiv:1409.2224}}}
  {[astro-ph.CO]}

\bibitem{Tansella:2017rpi}
V.~Tansella, C.~Bonvin, R.~Durrer, B.~Ghosh, E.~Sellentin, {The full-sky
  relativistic correlation function and power spectrum of galaxy number counts.
  Part I: theoretical aspects}.
\newblock JCAP \textbf{03}, 019 (2018).
\newblock \doi{10.1088/1475-7516/2018/03/019}.
\newblock {\href{https://arxiv.org/abs/1708.00492}{{arXiv:1708.00492}}}
  {[astro-ph.CO]}

\bibitem{Grimm:2020ays}
N.~Grimm, F.~Scaccabarozzi, J.~Yoo, S.G. Biern, J.O. Gong, {Galaxy Power
  Spectrum in General Relativity}.
\newblock JCAP \textbf{11}, 064 (2020).
\newblock \doi{10.1088/1475-7516/2020/11/064}.
\newblock {\href{https://arxiv.org/abs/2005.06484}{{arXiv:2005.06484}}}
  {[astro-ph.CO]}

\bibitem{Maartens:2012rh}
R.~Maartens, G.B. Zhao, D.~Bacon, K.~Koyama, A.~Raccanelli, {Relativistic
  corrections and non-Gaussianity in radio continuum surveys}.
\newblock JCAP \textbf{02}, 044 (2013).
\newblock \doi{10.1088/1475-7516/2013/02/044}.
\newblock {\href{https://arxiv.org/abs/1206.0732}{{arXiv:1206.0732}}}
  {[astro-ph.CO]}

\bibitem{Alonso:2015uua}
D.~Alonso, P.~Bull, P.G. Ferreira, R.~Maartens, M.~Santos, {Ultra large-scale
  cosmology in next-generation experiments with single tracers}.
\newblock Astrophys. J. \textbf{814}(2), 145 (2015).
\newblock \doi{10.1088/0004-637X/814/2/145}.
\newblock {\href{https://arxiv.org/abs/1505.07596}{{arXiv:1505.07596}}}
  {[astro-ph.CO]}

\bibitem{Alonso:2015sfa}
D.~Alonso, P.G. Ferreira, {Constraining ultralarge-scale cosmology with
  multiple tracers in optical and radio surveys}.
\newblock Phys. Rev. D \textbf{92}(6), 063,525 (2015).
\newblock \doi{10.1103/PhysRevD.92.063525}.
\newblock {\href{https://arxiv.org/abs/1507.03550}{{arXiv:1507.03550}}}
  {[astro-ph.CO]}

\bibitem{Fonseca:2015laa}
J.~Fonseca, S.~Camera, M.~Santos, R.~Maartens, {Hunting down horizon-scale
  effects with multi-wavelength surveys}.
\newblock Astrophys. J. Lett. \textbf{812}(2), L22 (2015).
\newblock \doi{10.1088/2041-8205/812/2/L22}.
\newblock {\href{https://arxiv.org/abs/1507.04605}{{arXiv:1507.04605}}}
  {[astro-ph.CO]}

\bibitem{Abramo:2017xnp}
L.R. Abramo, D.~Bertacca, {Disentangling the effects of Doppler velocity and
  primordial non-Gaussianity in galaxy power spectra}.
\newblock Phys. Rev. D \textbf{96}(12), 123,535 (2017).
\newblock \doi{10.1103/PhysRevD.96.123535}.
\newblock {\href{https://arxiv.org/abs/1706.01834}{{arXiv:1706.01834}}}
  {[astro-ph.CO]}

\bibitem{Barreira:2022sey}
A.~Barreira, {Can we actually constrain f$_{NL}$ using the scale-dependent bias
  effect? An illustration of the impact of galaxy bias uncertainties using the
  BOSS DR12 galaxy power spectrum}.
\newblock JCAP \textbf{11}, 013 (2022).
\newblock \doi{10.1088/1475-7516/2022/11/013}.
\newblock {\href{https://arxiv.org/abs/2205.05673}{{arXiv:2205.05673}}}
  {[astro-ph.CO]}

\bibitem{KiDS:2020suj}
M.~Asgari, et~al., {KiDS-1000 Cosmology: Cosmic shear constraints and
  comparison between two point statistics}.
\newblock Astron. Astrophys. \textbf{645}, A104 (2021).
\newblock \doi{10.1051/0004-6361/202039070}.
\newblock {\href{https://arxiv.org/abs/2007.15633}{{arXiv:2007.15633}}}
  {[astro-ph.CO]}

\bibitem{DES:2022ccp}
T.M.C. Abbott, et~al., {Dark Energy Survey Year 3 results: Constraints on
  extensions to \ensuremath{\Lambda}CDM with weak lensing and galaxy
  clustering}.
\newblock Phys. Rev. D \textbf{107}(8), 083,504 (2023).
\newblock \doi{10.1103/PhysRevD.107.083504}.
\newblock {\href{https://arxiv.org/abs/2207.05766}{{arXiv:2207.05766}}}
  {[astro-ph.CO]}

\bibitem{HSC:2018mrq}
C.~Hikage, et~al., {Cosmology from cosmic shear power spectra with Subaru Hyper
  Suprime-Cam first-year data}.
\newblock Publ. Astron. Soc. Jap. \textbf{71}(2), 43 (2019).
\newblock \doi{10.1093/pasj/psz010}.
\newblock {\href{https://arxiv.org/abs/1809.09148}{{arXiv:1809.09148}}}
  {[astro-ph.CO]}

\bibitem{Hamilton:1991es}
A.J.S. Hamilton, A.~Matthews, P.~Kumar, E.~Lu, {Reconstructing the primordial
  spectrum of fluctuations of the universe from the observed nonlinear
  clustering of galaxies}.
\newblock Astrophys. J. Lett. \textbf{374}, L1 (1991).
\newblock \doi{10.1086/186057}

\bibitem{Peacock:1996ci}
J.A. Peacock, S.J. Dodds, {Nonlinear evolution of cosmological power spectra}.
\newblock Mon. Not. Roy. Astron. Soc. \textbf{280}, L19 (1996).
\newblock \doi{10.1093/mnras/280.3.L19}.
\newblock
  {\href{https://arxiv.org/abs/astro-ph/9603031}{{arXiv:astro-ph/9603031}}}

\bibitem{Smith:2002dz}
R.E. Smith, J.A. Peacock, A.~Jenkins, S.D.M. White, C.S. Frenk, F.R. Pearce,
  P.A. Thomas, G.~Efstathiou, H.M.P. Couchmann, {Stable clustering, the halo
  model and nonlinear cosmological power spectra}.
\newblock Mon. Not. Roy. Astron. Soc. \textbf{341}, 1311 (2003).
\newblock \doi{10.1046/j.1365-8711.2003.06503.x}.
\newblock
  {\href{https://arxiv.org/abs/astro-ph/0207664}{{arXiv:astro-ph/0207664}}}

\bibitem{Heitmann:2009cu}
K.~Heitmann, D.~Higdon, M.~White, S.~Habib, B.J. Williams, C.~Wagner, {The
  Coyote Universe II: Cosmological Models and Precision Emulation of the
  Nonlinear Matter Power Spectrum}.
\newblock Astrophys. J. \textbf{705}, 156--174 (2009).
\newblock \doi{10.1088/0004-637X/705/1/156}.
\newblock {\href{https://arxiv.org/abs/0902.0429}{{arXiv:0902.0429}}}
  {[astro-ph.CO]}

\bibitem{Takahashi:2012em}
R.~Takahashi, M.~Sato, T.~Nishimichi, A.~Taruya, M.~Oguri, {Revising the
  Halofit Model for the Nonlinear Matter Power Spectrum}.
\newblock Astrophys. J. \textbf{761}, 152 (2012).
\newblock \doi{10.1088/0004-637X/761/2/152}.
\newblock {\href{https://arxiv.org/abs/1208.2701}{{arXiv:1208.2701}}}
  {[astro-ph.CO]}

\bibitem{Heitmann:2013bra}
K.~Heitmann, E.~Lawrence, J.~Kwan, S.~Habib, D.~Higdon, {The Coyote Universe
  Extended: Precision Emulation of the Matter Power Spectrum}.
\newblock Astrophys. J. \textbf{780}, 111 (2014).
\newblock \doi{10.1088/0004-637X/780/1/111}.
\newblock {\href{https://arxiv.org/abs/1304.7849}{{arXiv:1304.7849}}}
  {[astro-ph.CO]}

\bibitem{Mead:2015yca}
A.~Mead, J.~Peacock, C.~Heymans, S.~Joudaki, A.~Heavens, {An accurate halo
  model for fitting non-linear cosmological power spectra and baryonic feedback
  models}.
\newblock Mon. Not. Roy. Astron. Soc. \textbf{454}(2), 1958--1975 (2015).
\newblock \doi{10.1093/mnras/stv2036}.
\newblock {\href{https://arxiv.org/abs/1505.07833}{{arXiv:1505.07833}}}
  {[astro-ph.CO]}

\bibitem{Garrison:2017ssz}
L.H. Garrison, D.J. Eisenstein, D.~Ferrer, J.L. Tinker, P.A. Pinto, D.H.
  Weinberg, {The Abacus Cosmos: A Suite of Cosmological N-body Simulations}.
\newblock Astrophys. J. Suppl. \textbf{236}(2), 43 (2018).
\newblock \doi{10.3847/1538-4365/aabfd3}.
\newblock {\href{https://arxiv.org/abs/1712.05768}{{arXiv:1712.05768}}}
  {[astro-ph.CO]}

\bibitem{Bird:2018all}
S.~Bird, Y.~Ali-Ha\"\i{}moud, Y.~Feng, J.~Liu, {An Efficient and Accurate
  Hybrid Method for Simulating Non-Linear Neutrino Structure}.
\newblock Mon. Not. Roy. Astron. Soc. \textbf{481}(2), 1486--1500 (2018).
\newblock \doi{10.1093/mnras/sty2376}.
\newblock {\href{https://arxiv.org/abs/1803.09854}{{arXiv:1803.09854}}}
  {[astro-ph.CO]}

\bibitem{DeRose:2018xdj}
J.~DeRose, R.H. Wechsler, J.L. Tinker, M.R. Becker, Y.Y. Mao, T.~McClintock,
  S.~McLaughlin, E.~Rozo, Z.~Zhai, {The Aemulus Project I: Numerical
  Simulations for Precision Cosmology}.
\newblock Astrophys. J. \textbf{875}(1), 69 (2019).
\newblock \doi{10.3847/1538-4357/ab1085}.
\newblock {\href{https://arxiv.org/abs/1804.05865}{{arXiv:1804.05865}}}
  {[astro-ph.CO]}

\bibitem{Mead:2020vgs}
A.~Mead, S.~Brieden, T.~Tr\"oster, C.~Heymans, {HMcode-2020: Improved modelling
  of non-linear cosmological power spectra with baryonic feedback}  (2020).
\newblock \doi{10.1093/mnras/stab082}.
\newblock {\href{https://arxiv.org/abs/2009.01858}{{arXiv:2009.01858}}}
  {[astro-ph.CO]}

\bibitem{Linder:2003dr}
E.V. Linder, A.~Jenkins, {Cosmic structure and dark energy}.
\newblock Mon. Not. Roy. Astron. Soc. \textbf{346}, 573 (2003).
\newblock \doi{10.1046/j.1365-2966.2003.07112.x}.
\newblock
  {\href{https://arxiv.org/abs/astro-ph/0305286}{{arXiv:astro-ph/0305286}}}

\bibitem{Francis:2007qa}
M.J. Francis, G.F. Lewis, E.V. Linder, {Power Spectra to 1\% Accuracy between
  Dynamical Dark Energy Cosmologies}.
\newblock Mon. Not. Roy. Astron. Soc. \textbf{380}, 1079 (2007).
\newblock \doi{10.1111/j.1365-2966.2007.12139.x}.
\newblock {\href{https://arxiv.org/abs/0704.0312}{{arXiv:0704.0312}}}
  {[astro-ph]}

\bibitem{Casarini:2016ysv}
L.~Casarini, S.A. Bonometto, E.~Tessarotto, P.S. Corasaniti, {Extending the
  Coyote emulator to dark energy models with standard $w_0$-$w_a$
  parametrization of the equation of state}.
\newblock JCAP \textbf{08}, 008 (2016).
\newblock \doi{10.1088/1475-7516/2016/08/008}.
\newblock {\href{https://arxiv.org/abs/1601.07230}{{arXiv:1601.07230}}}
  {[astro-ph.CO]}

\bibitem{Lawrence:2017ost}
E.~Lawrence, K.~Heitmann, J.~Kwan, A.~Upadhye, D.~Bingham, S.~Habib, D.~Higdon,
  A.~Pope, H.~Finkel, N.~Frontiere, {The Mira-Titan Universe II: Matter Power
  Spectrum Emulation}.
\newblock Astrophys. J. \textbf{847}(1), 50 (2017).
\newblock \doi{10.3847/1538-4357/aa86a9}.
\newblock {\href{https://arxiv.org/abs/1705.03388}{{arXiv:1705.03388}}}
  {[astro-ph.CO]}

\bibitem{Euclid:2020rfv}
M.~Knabenhans, et~al., {Euclid preparation: IX. EuclidEmulator2 \textendash{}
  power spectrum emulation with massive neutrinos and self-consistent dark
  energy perturbations}.
\newblock Mon. Not. Roy. Astron. Soc. \textbf{505}(2), 2840--2869 (2021).
\newblock \doi{10.1093/mnras/stab1366}.
\newblock {\href{https://arxiv.org/abs/2010.11288}{{arXiv:2010.11288}}}
  {[astro-ph.CO]}

\bibitem{Stabenau:2006td}
H.F. Stabenau, B.~Jain, {N-Body Simulations of Alternate Gravity Models}.
\newblock Phys. Rev. D \textbf{74}, 084,007 (2006).
\newblock \doi{10.1103/PhysRevD.74.084007}.
\newblock
  {\href{https://arxiv.org/abs/astro-ph/0604038}{{arXiv:astro-ph/0604038}}}

\bibitem{Laszlo:2007td}
I.~Laszlo, R.~Bean, {Nonlinear growth in modified gravity theories of dark
  energy}.
\newblock Phys. Rev. D \textbf{77}, 024,048 (2008).
\newblock \doi{10.1103/PhysRevD.77.024048}.
\newblock {\href{https://arxiv.org/abs/0709.0307}{{arXiv:0709.0307}}}
  {[astro-ph]}

\bibitem{Oyaizu:2008tb}
H.~Oyaizu, M.~Lima, W.~Hu, {Nonlinear evolution of f(R) cosmologies. 2. Power
  spectrum}.
\newblock Phys. Rev. D \textbf{78}, 123,524 (2008).
\newblock \doi{10.1103/PhysRevD.78.123524}.
\newblock {\href{https://arxiv.org/abs/0807.2462}{{arXiv:0807.2462}}}
  {[astro-ph]}

\bibitem{Baldi:2008ay}
M.~Baldi, V.~Pettorino, G.~Robbers, V.~Springel, {Hydrodynamical N-body
  simulations of coupled dark energy cosmologies}.
\newblock Mon. Not. Roy. Astron. Soc. \textbf{403}, 1684--1702 (2010).
\newblock \doi{10.1111/j.1365-2966.2009.15987.x}.
\newblock {\href{https://arxiv.org/abs/0812.3901}{{arXiv:0812.3901}}}
  {[astro-ph]}

\bibitem{Koyama:2009me}
K.~Koyama, A.~Taruya, T.~Hiramatsu, {Non-linear Evolution of Matter Power
  Spectrum in Modified Theory of Gravity}.
\newblock Phys. Rev. D \textbf{79}, 123,512 (2009).
\newblock \doi{10.1103/PhysRevD.79.123512}.
\newblock {\href{https://arxiv.org/abs/0902.0618}{{arXiv:0902.0618}}}
  {[astro-ph.CO]}

\bibitem{Schmidt:2009sg}
F.~Schmidt, {Self-Consistent Cosmological Simulations of DGP Braneworld
  Gravity}.
\newblock Phys. Rev. D \textbf{80}, 043,001 (2009).
\newblock \doi{10.1103/PhysRevD.80.043001}.
\newblock {\href{https://arxiv.org/abs/0905.0858}{{arXiv:0905.0858}}}
  {[astro-ph.CO]}

\bibitem{Chan:2009ew}
K.C. Chan, R.~Scoccimarro, {Large-Scale Structure in Brane-Induced Gravity II.
  Numerical Simulations}.
\newblock Phys. Rev. D \textbf{80}, 104,005 (2009).
\newblock \doi{10.1103/PhysRevD.80.104005}.
\newblock {\href{https://arxiv.org/abs/0906.4548}{{arXiv:0906.4548}}}
  {[astro-ph.CO]}

\bibitem{Zhao:2010qy}
G.B. Zhao, B.~Li, K.~Koyama, {N-body Simulations for f(R) Gravity using a
  Self-adaptive Particle-Mesh Code}.
\newblock Phys. Rev. D \textbf{83}, 044,007 (2011).
\newblock \doi{10.1103/PhysRevD.83.044007}.
\newblock {\href{https://arxiv.org/abs/1011.1257}{{arXiv:1011.1257}}}
  {[astro-ph.CO]}

\bibitem{Li:2013tda}
B.~Li, A.~Barreira, C.M. Baugh, W.A. Hellwing, K.~Koyama, S.~Pascoli, G.B.
  Zhao, {Simulating the quartic Galileon gravity model on adaptively refined
  meshes}.
\newblock JCAP \textbf{11}, 012 (2013).
\newblock \doi{10.1088/1475-7516/2013/11/012}.
\newblock {\href{https://arxiv.org/abs/1308.3491}{{arXiv:1308.3491}}}
  {[astro-ph.CO]}

\bibitem{Barreira:2014kra}
A.~Barreira, B.~Li, W.A. Hellwing, C.M. Baugh, S.~Pascoli, {Nonlinear structure
  formation in Nonlocal Gravity}.
\newblock JCAP \textbf{09}, 031 (2014).
\newblock \doi{10.1088/1475-7516/2014/09/031}.
\newblock {\href{https://arxiv.org/abs/1408.1084}{{arXiv:1408.1084}}}
  {[astro-ph.CO]}

\bibitem{Winther:2015wla}
H.A. Winther, et~al., {Modified Gravity N-body Code Comparison Project}.
\newblock Mon. Not. Roy. Astron. Soc. \textbf{454}(4), 4208--4234 (2015).
\newblock \doi{10.1093/mnras/stv2253}.
\newblock {\href{https://arxiv.org/abs/1506.06384}{{arXiv:1506.06384}}}
  {[astro-ph.CO]}

\bibitem{Bose:2016wms}
S.~Bose, B.~Li, A.~Barreira, J.h. He, W.A. Hellwing, K.~Koyama, C.~Llinares,
  G.B. Zhao, {Speeding up $N$-body simulations of modified gravity: Chameleon
  screening models}.
\newblock JCAP \textbf{02}, 050 (2017).
\newblock \doi{10.1088/1475-7516/2017/02/050}.
\newblock {\href{https://arxiv.org/abs/1611.09375}{{arXiv:1611.09375}}}
  {[astro-ph.CO]}

\bibitem{Lesgourgues:2006nd}
J.~Lesgourgues, S.~Pastor, {Massive neutrinos and cosmology}.
\newblock Phys. Rept. \textbf{429}, 307--379 (2006).
\newblock \doi{10.1016/j.physrep.2006.04.001}.
\newblock
  {\href{https://arxiv.org/abs/astro-ph/0603494}{{arXiv:astro-ph/0603494}}}

\bibitem{Newman:2022rbn}
J.A. Newman, D.~Gruen, {Photometric Redshifts for Next-Generation Surveys}.
\newblock Ann. Rev. Astron. Astrophys. \textbf{60}, 363--414 (2022).
\newblock \doi{10.1146/annurev-astro-032122-014611}.
\newblock {\href{https://arxiv.org/abs/2206.13633}{{arXiv:2206.13633}}}
  {[astro-ph.CO]}

\bibitem{DES:2021vln}
L.F. Secco, et~al., {Dark Energy Survey Year 3 results: Cosmology from cosmic
  shear and robustness to modeling uncertainty}.
\newblock Phys. Rev. D \textbf{105}(2), 023,515 (2022).
\newblock \doi{10.1103/PhysRevD.105.023515}.
\newblock {\href{https://arxiv.org/abs/2105.13544}{{arXiv:2105.13544}}}
  {[astro-ph.CO]}

\bibitem{Bartelmann:1999yn}
M.~Bartelmann, P.~Schneider, {Weak gravitational lensing}.
\newblock Phys. Rept. \textbf{340}, 291--472 (2001).
\newblock \doi{10.1016/S0370-1573(00)00082-X}.
\newblock
  {\href{https://arxiv.org/abs/astro-ph/9912508}{{arXiv:astro-ph/9912508}}}

\bibitem{Hoekstra:2008db}
H.~Hoekstra, B.~Jain, {Weak Gravitational Lensing and its Cosmological
  Applications}.
\newblock Ann. Rev. Nucl. Part. Sci. \textbf{58}, 99--123 (2008).
\newblock \doi{10.1146/annurev.nucl.58.110707.171151}.
\newblock {\href{https://arxiv.org/abs/0805.0139}{{arXiv:0805.0139}}}
  {[astro-ph]}

\bibitem{Huterer_book}
D.~Huterer, \emph{A Course in Cosmology: From Theory to Practice} (Cambridge
  University Press, Cambridge, 2023)

\bibitem{Hu:2002rm}
W.~Hu, {Dark energy and matter evolution from lensing tomography}.
\newblock Phys. Rev. D \textbf{66}, 083,515 (2002).
\newblock \doi{10.1103/PhysRevD.66.083515}.
\newblock
  {\href{https://arxiv.org/abs/astro-ph/0208093}{{arXiv:astro-ph/0208093}}}

\bibitem{Catelan:2000vm}
P.~Catelan, M.~Kamionkowski, R.D. Blandford, {Intrinsic and extrinsic galaxy
  alignment}.
\newblock Mon. Not. Roy. Astron. Soc. \textbf{320}, L7--L13 (2001).
\newblock \doi{10.1046/j.1365-8711.2001.04105.x}.
\newblock
  {\href{https://arxiv.org/abs/astro-ph/0005470}{{arXiv:astro-ph/0005470}}}

\bibitem{Hirata:2004gc}
C.M. Hirata, U.~Seljak, {Intrinsic alignment-lensing interference as a
  contaminant of cosmic shear}.
\newblock Phys. Rev. D \textbf{70}, 063,526 (2004).
\newblock \doi{10.1103/PhysRevD.82.049901}.
\newblock [Erratum: Phys.Rev.D 82, 049901 (2010)].
\newblock
  {\href{https://arxiv.org/abs/astro-ph/0406275}{{arXiv:astro-ph/0406275}}}

\bibitem{Mandelbaum:2017jpr}
R.~Mandelbaum, {Weak lensing for precision cosmology}.
\newblock Ann. Rev. Astron. Astrophys. \textbf{56}, 393--433 (2018).
\newblock \doi{10.1146/annurev-astro-081817-051928}.
\newblock {\href{https://arxiv.org/abs/1710.03235}{{arXiv:1710.03235}}}
  {[astro-ph.CO]}

\bibitem{DES:2021wwk}
T.M.C. Abbott, et~al., {Dark Energy Survey Year 3 results: Cosmological
  constraints from galaxy clustering and weak lensing}.
\newblock Phys. Rev. D \textbf{105}(2), 023,520 (2022).
\newblock \doi{10.1103/PhysRevD.105.023520}.
\newblock {\href{https://arxiv.org/abs/2105.13549}{{arXiv:2105.13549}}}
  {[astro-ph.CO]}

\bibitem{Bond:1990iw}
J.R. Bond, S.~Cole, G.~Efstathiou, N.~Kaiser, {Excursion set mass functions for
  hierarchical Gaussian fluctuations}.
\newblock Astrophys. J. \textbf{379}, 440 (1991).
\newblock \doi{10.1086/170520}

\bibitem{Zentner:2006vw}
A.R. Zentner, {The Excursion Set Theory of Halo Mass Functions, Halo
  Clustering, and Halo Growth}.
\newblock Int. J. Mod. Phys. D \textbf{16}, 763--816 (2007).
\newblock \doi{10.1142/S0218271807010511}.
\newblock
  {\href{https://arxiv.org/abs/astro-ph/0611454}{{arXiv:astro-ph/0611454}}}

\bibitem{Press:1973iz}
W.H. Press, P.~Schechter, {Formation of galaxies and clusters of galaxies by
  selfsimilar gravitational condensation}.
\newblock Astrophys. J. \textbf{187}, 425--438 (1974).
\newblock \doi{10.1086/152650}

\bibitem{Allen:2011zs}
S.W. Allen, A.E. Evrard, A.B. Mantz, {Cosmological Parameters from Observations
  of Galaxy Clusters}.
\newblock Ann. Rev. Astron. Astrophys. \textbf{49}, 409--470 (2011).
\newblock \doi{10.1146/annurev-astro-081710-102514}.
\newblock {\href{https://arxiv.org/abs/1103.4829}{{arXiv:1103.4829}}}
  {[astro-ph.CO]}

\bibitem{Haiman:2000bw}
Z.~Haiman, J.J. Mohr, G.P. Holder, {Constraints on quintessence from future
  galaxy cluster surveys}.
\newblock Astrophys. J. \textbf{553}, 545 (2000).
\newblock \doi{10.1086/320939}.
\newblock
  {\href{https://arxiv.org/abs/astro-ph/0002336}{{arXiv:astro-ph/0002336}}}

\bibitem{Tinker:2008ff}
J.L. Tinker, A.V. Kravtsov, A.~Klypin, K.~Abazajian, M.S. Warren, G.~Yepes,
  S.~Gottlober, D.E. Holz, {Toward a halo mass function for precision
  cosmology: The Limits of universality}.
\newblock Astrophys. J. \textbf{688}, 709--728 (2008).
\newblock \doi{10.1086/591439}.
\newblock {\href{https://arxiv.org/abs/0803.2706}{{arXiv:0803.2706}}}
  {[astro-ph]}

\bibitem{McClintock:2018uyf}
T.~McClintock, E.~Rozo, M.R. Becker, J.~DeRose, Y.Y. Mao, S.~McLaughlin, J.L.
  Tinker, R.H. Wechsler, Z.~Zhai, {The Aemulus Project II: Emulating the Halo
  Mass Function}.
\newblock Astrophys. J. \textbf{872}(1), 53 (2019).
\newblock \doi{10.3847/1538-4357/aaf568}.
\newblock {\href{https://arxiv.org/abs/1804.05866}{{arXiv:1804.05866}}}
  {[astro-ph.CO]}

\bibitem{Nishimichi:2018etk}
T.~Nishimichi, et~al., {Dark Quest. I. Fast and Accurate Emulation of Halo
  Clustering Statistics and Its Application to Galaxy Clustering}.
\newblock Astrophys. J. \textbf{884}, 29 (2019).
\newblock \doi{10.3847/1538-4357/ab3719}.
\newblock {\href{https://arxiv.org/abs/1811.09504}{{arXiv:1811.09504}}}
  {[astro-ph.CO]}

\bibitem{Bocquet:2020tes}
S.~Bocquet, K.~Heitmann, S.~Habib, E.~Lawrence, T.~Uram, N.~Frontiere, A.~Pope,
  H.~Finkel, {The Mira-Titan Universe. III. Emulation of the Halo Mass
  Function}.
\newblock Astrophys. J. \textbf{901}(1), 5 (2020).
\newblock \doi{10.3847/1538-4357/abac5c}.
\newblock {\href{https://arxiv.org/abs/2003.12116}{{arXiv:2003.12116}}}
  {[astro-ph.CO]}

\bibitem{Bahcall:1998ur}
N.A. Bahcall, X.h. Fan, {The Most Massive Distant Clusters: Determining
  $\Omega$ and $\sigma_8$}.
\newblock Astrophys. J. \textbf{504}, 1 (1998).
\newblock \doi{10.1086/306088}.
\newblock
  {\href{https://arxiv.org/abs/astro-ph/9803277}{{arXiv:astro-ph/9803277}}}

\bibitem{Mullis:2005hp}
C.R. Mullis, et~al., {Discovery of an X-ray-Luminous Galaxy Cluster at z=1.4}.
\newblock Astrophys. J. \textbf{623}, L85--L88 (2005).
\newblock \doi{10.1086/429801}.
\newblock
  {\href{https://arxiv.org/abs/astro-ph/0503004}{{arXiv:astro-ph/0503004}}}

\bibitem{Brodwin:2010ig}
M.~Brodwin, et~al., {SPT-CL J0546-5345: A Massive z > 1 Galaxy Cluster Selected
  Via the Sunyaev-Zel'dovich Effect with the South Pole Telescope}.
\newblock Astrophys. J. \textbf{721}, 90--97 (2010).
\newblock \doi{10.1088/0004-637X/721/1/90}.
\newblock {\href{https://arxiv.org/abs/1006.5639}{{arXiv:1006.5639}}}

\bibitem{Holz:2010ck}
D.E. Holz, S.~Perlmutter, {The most massive objects in the Universe}.
\newblock Astrophys. J. Lett. \textbf{755}, L36 (2012).
\newblock \doi{10.1088/2041-8205/755/2/L36}.
\newblock {\href{https://arxiv.org/abs/1004.5349}{{arXiv:1004.5349}}}
  {[astro-ph.CO]}

\bibitem{Kravtsov:2012zs}
A.~Kravtsov, S.~Borgani, {Formation of Galaxy Clusters}.
\newblock Ann. Rev. Astron. Astrophys. \textbf{50}, 353--409 (2012).
\newblock \doi{10.1146/annurev-astro-081811-125502}.
\newblock {\href{https://arxiv.org/abs/1205.5556}{{arXiv:1205.5556}}}
  {[astro-ph.CO]}

\bibitem{Mortonson:2010mj}
M.J. Mortonson, W.~Hu, D.~Huterer, {Simultaneous Falsification of $\Lambda$CDM
  and Quintessence with Massive, Distant Clusters}.
\newblock Phys. Rev. D \textbf{83}, 023,015 (2011).
\newblock \doi{10.1103/PhysRevD.83.023015}.
\newblock {\href{https://arxiv.org/abs/1011.0004}{{arXiv:1011.0004}}}
  {[astro-ph.CO]}

\bibitem{Jee:2013gey}
M.J. Jee, J.P. Hughes, F.~Menanteau, C.~Sifon, R.~Mandelbaum, L.F. Barrientos,
  L.~Infante, K.Y. Ng, {Weighing ''El Gordo'' with a Precision Scale: Hubble
  Space Telescope Weak-lensing Analysis of the Merging Galaxy Cluster ACT-CL
  J0102\textendash{}4915 at $z$ = 0.87}.
\newblock Astrophys. J. \textbf{785}, 20 (2014).
\newblock \doi{10.1088/0004-637X/785/1/20}.
\newblock {\href{https://arxiv.org/abs/1309.5097}{{arXiv:1309.5097}}}
  {[astro-ph.CO]}

\bibitem{Song:2008qt}
Y.S. Song, W.J. Percival, {Reconstructing the history of structure formation
  using Redshift Distortions}.
\newblock JCAP \textbf{10}, 004 (2009).
\newblock \doi{10.1088/1475-7516/2009/10/004}.
\newblock {\href{https://arxiv.org/abs/0807.0810}{{arXiv:0807.0810}}}
  {[astro-ph]}

\bibitem{delaTorre:2012dg}
S.~de~la Torre, L.~Guzzo, {Modelling non-linear redshift-space distortions in
  the galaxy clustering pattern: systematic errors on the growth rate
  parameter}.
\newblock Mon. Not. Roy. Astron. Soc. \textbf{427}, 327 (2012).
\newblock \doi{10.1111/j.1365-2966.2012.21824.x}.
\newblock {\href{https://arxiv.org/abs/1202.5559}{{arXiv:1202.5559}}}
  {[astro-ph.CO]}

\bibitem{Howlett:2022len}
C.~Howlett, K.~Said, J.R. Lucey, M.~Colless, F.~Qin, Y.~Lai, R.B. Tully, T.M.
  Davis, {The sloan digital sky survey peculiar velocity catalogue}.
\newblock Mon. Not. Roy. Astron. Soc. \textbf{515}(1), 953--976 (2022).
\newblock \doi{10.1093/mnras/stac1681}.
\newblock {\href{https://arxiv.org/abs/2201.03112}{{arXiv:2201.03112}}}
  {[astro-ph.CO]}

\bibitem{Tully:2022rbj}
R.B. Tully, et~al., {Cosmicflows-4}.
\newblock Astrophys. J. \textbf{944}(1), 94 (2023).
\newblock \doi{10.3847/1538-4357/ac94d8}.
\newblock {\href{https://arxiv.org/abs/2209.11238}{{arXiv:2209.11238}}}
  {[astro-ph.CO]}

\bibitem{Hand:2012ui}
N.~Hand, et~al., {Evidence of Galaxy Cluster Motions with the Kinematic
  Sunyaev-Zel'dovich Effect}.
\newblock Phys. Rev. Lett. \textbf{109}, 041,101 (2012).
\newblock \doi{10.1103/PhysRevLett.109.041101}.
\newblock {\href{https://arxiv.org/abs/1203.4219}{{arXiv:1203.4219}}}
  {[astro-ph.CO]}

\bibitem{Peacock:2018xlz}
J.A. Peacock, M.~Bilicki, {Wide-area tomography of CMB lensing and the growth
  of cosmological density fluctuations}.
\newblock Mon. Not. Roy. Astron. Soc. \textbf{481}(1), 1133--1148 (2018).
\newblock \doi{10.1093/mnras/sty2314}.
\newblock {\href{https://arxiv.org/abs/1805.11525}{{arXiv:1805.11525}}}
  {[astro-ph.CO]}

\bibitem{Wilson:2019brt}
M.J. Wilson, M.~White, {Cosmology with dropout selection: straw-man surveys
  \textbackslash{}\& CMB lensing}.
\newblock JCAP \textbf{10}, 015 (2019).
\newblock \doi{10.1088/1475-7516/2019/10/015}.
\newblock {\href{https://arxiv.org/abs/1904.13378}{{arXiv:1904.13378}}}
  {[astro-ph.CO]}

\bibitem{Krolewski:2019yrv}
A.~Krolewski, S.~Ferraro, E.F. Schlafly, M.~White, {unWISE tomography of Planck
  CMB lensing}.
\newblock JCAP \textbf{05}, 047 (2020).
\newblock \doi{10.1088/1475-7516/2020/05/047}.
\newblock {\href{https://arxiv.org/abs/1909.07412}{{arXiv:1909.07412}}}
  {[astro-ph.CO]}

\bibitem{Garcia-Garcia:2021unp}
C.~Garc\'\i{}a-Garc\'\i{}a, J.R. Zapatero, D.~Alonso, E.~Bellini, P.G.
  Ferreira, E.M. Mueller, A.~Nicola, P.~Ruiz-Lapuente, {The growth of density
  perturbations in the last \ensuremath{\sim}10 billion years from tomographic
  large-scale structure data}.
\newblock JCAP \textbf{10}, 030 (2021).
\newblock \doi{10.1088/1475-7516/2021/10/030}.
\newblock {\href{https://arxiv.org/abs/2105.12108}{{arXiv:2105.12108}}}
  {[astro-ph.CO]}

\bibitem{Boruah:2019icj}
S.S. Boruah, M.J. Hudson, G.~Lavaux, {Cosmic flows in the nearby Universe: new
  peculiar velocities from SNe and cosmological constraints}.
\newblock Mon. Not. Roy. Astron. Soc. \textbf{498}(2), 2703--2718 (2020).
\newblock \doi{10.1093/mnras/staa2485}.
\newblock {\href{https://arxiv.org/abs/1912.09383}{{arXiv:1912.09383}}}
  {[astro-ph.CO]}

\bibitem{Huterer:2016uyq}
D.~Huterer, D.~Shafer, D.~Scolnic, F.~Schmidt, {Testing $\Lambda$CDM at the
  lowest redshifts with SN Ia and galaxy velocities}.
\newblock JCAP \textbf{05}, 015 (2017).
\newblock \doi{10.1088/1475-7516/2017/05/015}.
\newblock {\href{https://arxiv.org/abs/1611.09862}{{arXiv:1611.09862}}}
  {[astro-ph.CO]}

\bibitem{Turner:2022mla}
R.J. Turner, C.~Blake, R.~Ruggeri, {A local measurement of the growth rate from
  peculiar velocities and galaxy clustering correlations in the 6dF Galaxy
  Survey}.
\newblock Mon. Not. Roy. Astron. Soc. \textbf{518}(2), 2436--2452 (2022).
\newblock \doi{10.1093/mnras/stac3256}.
\newblock {\href{https://arxiv.org/abs/2207.03707}{{arXiv:2207.03707}}}
  {[astro-ph.CO]}

\bibitem{Said:2020epb}
K.~Said, M.~Colless, C.~Magoulas, J.R. Lucey, M.J. Hudson, {Joint analysis of
  6dFGS and SDSS peculiar velocities for the growth rate of cosmic structure
  and tests of gravity}.
\newblock Mon. Not. Roy. Astron. Soc. \textbf{497}(1), 1275--1293 (2020).
\newblock \doi{10.1093/mnras/staa2032}.
\newblock {\href{https://arxiv.org/abs/2007.04993}{{arXiv:2007.04993}}}
  {[astro-ph.CO]}

\bibitem{Beutler:2012px}
F.~Beutler, C.~Blake, M.~Colless, D.H. Jones, L.~Staveley-Smith, G.B. Poole,
  L.~Campbell, Q.~Parker, W.~Saunders, F.~Watson, {The 6dF Galaxy Survey: $z
  \approx 0$ measurement of the growth rate and $\sigma_8$}.
\newblock Mon. Not. Roy. Astron. Soc. \textbf{423}, 3430--3444 (2012).
\newblock \doi{10.1111/j.1365-2966.2012.21136.x}.
\newblock {\href{https://arxiv.org/abs/1204.4725}{{arXiv:1204.4725}}}
  {[astro-ph.CO]}

\bibitem{Blake:2013nif}
C.~Blake, et~al., {Galaxy And Mass Assembly (GAMA): improved cosmic growth
  measurements using multiple tracers of large-scale structure}.
\newblock Mon. Not. Roy. Astron. Soc. \textbf{436}, 3089 (2013).
\newblock \doi{10.1093/mnras/stt1791}.
\newblock {\href{https://arxiv.org/abs/1309.5556}{{arXiv:1309.5556}}}
  {[astro-ph.CO]}

\bibitem{Blake:2012pj}
C.~Blake, et~al., {The WiggleZ Dark Energy Survey: Joint measurements of the
  expansion and growth history at z \ensuremath{<} 1}.
\newblock Mon. Not. Roy. Astron. Soc. \textbf{425}, 405--414 (2012).
\newblock \doi{10.1111/j.1365-2966.2012.21473.x}.
\newblock {\href{https://arxiv.org/abs/1204.3674}{{arXiv:1204.3674}}}
  {[astro-ph.CO]}

\bibitem{Pezzotta:2016gbo}
A.~Pezzotta, et~al., {The VIMOS Public Extragalactic Redshift Survey (VIPERS):
  The growth of structure at $0.5 < z < 1.2$ from redshift-space distortions in
  the clustering of the PDR-2 final sample}.
\newblock Astron. Astrophys. \textbf{604}, A33 (2017).
\newblock \doi{10.1051/0004-6361/201630295}.
\newblock {\href{https://arxiv.org/abs/1612.05645}{{arXiv:1612.05645}}}
  {[astro-ph.CO]}

\bibitem{Howlett:2014opa}
C.~Howlett, A.~Ross, L.~Samushia, W.~Percival, M.~Manera, {The clustering of
  the SDSS main galaxy sample \textendash{} II. Mock galaxy catalogues and a
  measurement of the growth of structure from redshift space distortions at $z
  = 0.15$}.
\newblock Mon. Not. Roy. Astron. Soc. \textbf{449}(1), 848--866 (2015).
\newblock \doi{10.1093/mnras/stu2693}.
\newblock {\href{https://arxiv.org/abs/1409.3238}{{arXiv:1409.3238}}}
  {[astro-ph.CO]}

\bibitem{Okumura:2015lvp}
T.~Okumura, et~al., {The Subaru FMOS galaxy redshift survey (FastSound). IV.
  New constraint on gravity theory from redshift space distortions at $z\sim
  1.4$}.
\newblock Publ. Astron. Soc. Jap. \textbf{68}(3), 38 (2016).
\newblock \doi{10.1093/pasj/psw029}.
\newblock {\href{https://arxiv.org/abs/1511.08083}{{arXiv:1511.08083}}}
  {[astro-ph.CO]}

\bibitem{eBOSS:2020yzd}
S.~Alam, et~al., {Completed SDSS-IV extended Baryon Oscillation Spectroscopic
  Survey: Cosmological implications from two decades of spectroscopic surveys
  at the Apache Point Observatory}.
\newblock Phys. Rev. D \textbf{103}(8), 083,533 (2021).
\newblock \doi{10.1103/PhysRevD.103.083533}.
\newblock {\href{https://arxiv.org/abs/2007.08991}{{arXiv:2007.08991}}}
  {[astro-ph.CO]}

\bibitem{Planck:2018vyg}
N.~Aghanim, et~al., {Planck 2018 results. VI. Cosmological parameters}.
\newblock Astron. Astrophys. \textbf{641}, A6 (2020).
\newblock \doi{10.1051/0004-6361/201833910}.
\newblock [Erratum: Astron.Astrophys. 652, C4 (2021)].
\newblock {\href{https://arxiv.org/abs/1807.06209}{{arXiv:1807.06209}}}
  {[astro-ph.CO]}

\bibitem{ACT:2020gnv}
S.~Aiola, et~al., {The Atacama Cosmology Telescope: DR4 Maps and Cosmological
  Parameters}.
\newblock JCAP \textbf{12}, 047 (2020).
\newblock \doi{10.1088/1475-7516/2020/12/047}.
\newblock {\href{https://arxiv.org/abs/2007.07288}{{arXiv:2007.07288}}}
  {[astro-ph.CO]}

\bibitem{DES:2021bvc}
A.~Amon, et~al., {Dark Energy Survey Year 3 results: Cosmology from cosmic
  shear and robustness to data calibration}.
\newblock Phys. Rev. D \textbf{105}(2), 023,514 (2022).
\newblock \doi{10.1103/PhysRevD.105.023514}.
\newblock {\href{https://arxiv.org/abs/2105.13543}{{arXiv:2105.13543}}}
  {[astro-ph.CO]}

\bibitem{DES:2022urg}
T.M.C. Abbott, et~al., {Joint analysis of Dark Energy Survey Year 3 data and
  CMB lensing from SPT and Planck. III. Combined cosmological constraints}.
\newblock Phys. Rev. D \textbf{107}(2), 023,531 (2023).
\newblock \doi{10.1103/PhysRevD.107.023531}.
\newblock {\href{https://arxiv.org/abs/2206.10824}{{arXiv:2206.10824}}}
  {[astro-ph.CO]}

\bibitem{Heymans:2020gsg}
C.~Heymans, et~al., {KiDS-1000 Cosmology: Multi-probe weak gravitational
  lensing and spectroscopic galaxy clustering constraints}.
\newblock Astron. Astrophys. \textbf{646}, A140 (2021).
\newblock \doi{10.1051/0004-6361/202039063}.
\newblock {\href{https://arxiv.org/abs/2007.15632}{{arXiv:2007.15632}}}
  {[astro-ph.CO]}

\bibitem{Sugiyama:2023fzm}
S.~Sugiyama, et~al., {Hyper Suprime-Cam Year 3 Results: Cosmology from Galaxy
  Clustering and Weak Lensing with HSC and SDSS using the Minimal Bias Model}
  (2023).
\newblock {\href{https://arxiv.org/abs/2304.00705}{{arXiv:2304.00705}}}
  {[astro-ph.CO]}

\bibitem{Li:2023tui}
X.~Li, et~al., {Hyper Suprime-Cam Year 3 Results: Cosmology from Cosmic Shear
  Two-point Correlation Functions}  (2023).
\newblock {\href{https://arxiv.org/abs/2304.00702}{{arXiv:2304.00702}}}
  {[astro-ph.CO]}

\bibitem{Dalal:2023olq}
R.~Dalal, et~al., {Hyper Suprime-Cam Year 3 Results: Cosmology from Cosmic
  Shear Power Spectra}  (2023).
\newblock {\href{https://arxiv.org/abs/2304.00701}{{arXiv:2304.00701}}}
  {[astro-ph.CO]}

\bibitem{Kobayashi:2021oud}
Y.~Kobayashi, T.~Nishimichi, M.~Takada, H.~Miyatake, {Full-shape cosmology
  analysis of the SDSS-III BOSS galaxy power spectrum using an emulator-based
  halo model: A 5\% determination of \ensuremath{\sigma}8}.
\newblock Phys. Rev. D \textbf{105}(8), 083,517 (2022).
\newblock \doi{10.1103/PhysRevD.105.083517}.
\newblock {\href{https://arxiv.org/abs/2110.06969}{{arXiv:2110.06969}}}
  {[astro-ph.CO]}

\bibitem{Philcox:2021kcw}
O.H.E. Philcox, M.M. Ivanov, {BOSS DR12 full-shape cosmology:
  \ensuremath{\Lambda}CDM constraints from the large-scale galaxy power
  spectrum and bispectrum monopole}.
\newblock Phys. Rev. D \textbf{105}(4), 043,517 (2022).
\newblock \doi{10.1103/PhysRevD.105.043517}.
\newblock {\href{https://arxiv.org/abs/2112.04515}{{arXiv:2112.04515}}}
  {[astro-ph.CO]}

\bibitem{Chen:2022jzq}
S.F. Chen, M.~White, J.~DeRose, N.~Kokron, {Cosmological analysis of
  three-dimensional BOSS galaxy clustering and Planck CMB lensing cross
  correlations via Lagrangian perturbation theory}.
\newblock JCAP \textbf{07}(07), 041 (2022).
\newblock \doi{10.1088/1475-7516/2022/07/041}.
\newblock {\href{https://arxiv.org/abs/2204.10392}{{arXiv:2204.10392}}}
  {[astro-ph.CO]}

\bibitem{Krolewski:2021yqy}
A.~Krolewski, S.~Ferraro, M.~White, {Cosmological constraints from unWISE and
  Planck CMB lensing tomography}.
\newblock JCAP \textbf{12}(12), 028 (2021).
\newblock \doi{10.1088/1475-7516/2021/12/028}.
\newblock {\href{https://arxiv.org/abs/2105.03421}{{arXiv:2105.03421}}}
  {[astro-ph.CO]}

\bibitem{White:2021yvw}
M.~White, et~al., {Cosmological constraints from the tomographic
  cross-correlation of DESI Luminous Red Galaxies and Planck CMB lensing}.
\newblock JCAP \textbf{02}(02), 007 (2022).
\newblock \doi{10.1088/1475-7516/2022/02/007}.
\newblock {\href{https://arxiv.org/abs/2111.09898}{{arXiv:2111.09898}}}
  {[astro-ph.CO]}

\bibitem{SPT:2018njh}
S.~Bocquet, et~al., {Cluster Cosmology Constraints from the 2500 deg$^2$ SPT-SZ
  Survey: Inclusion of Weak Gravitational Lensing Data from Magellan and the
  Hubble Space Telescope}.
\newblock Astrophys. J. \textbf{878}(1), 55 (2019).
\newblock \doi{10.3847/1538-4357/ab1f10}.
\newblock {\href{https://arxiv.org/abs/1812.01679}{{arXiv:1812.01679}}}
  {[astro-ph.CO]}

\bibitem{DES:2020ahh}
T.M.C. Abbott, et~al., {Dark Energy Survey Year 1 Results: Cosmological
  constraints from cluster abundances and weak lensing}.
\newblock Phys. Rev. D \textbf{102}(2), 023,509 (2020).
\newblock \doi{10.1103/PhysRevD.102.023509}.
\newblock {\href{https://arxiv.org/abs/2002.11124}{{arXiv:2002.11124}}}
  {[astro-ph.CO]}

\bibitem{Lue:2003ky}
A.~Lue, R.~Scoccimarro, G.~Starkman, {Differentiating between modified gravity
  and dark energy}.
\newblock Phys. Rev. D \textbf{69}, 044,005 (2004).
\newblock \doi{10.1103/PhysRevD.69.044005}.
\newblock
  {\href{https://arxiv.org/abs/astro-ph/0307034}{{arXiv:astro-ph/0307034}}}

\bibitem{Amendola:2007rr}
L.~Amendola, M.~Kunz, D.~Sapone, {Measuring the dark side (with weak lensing)}.
\newblock JCAP \textbf{04}, 013 (2008).
\newblock \doi{10.1088/1475-7516/2008/04/013}.
\newblock {\href{https://arxiv.org/abs/0704.2421}{{arXiv:0704.2421}}}
  {[astro-ph]}

\bibitem{Linder:2005in}
E.V. Linder, {Cosmic growth history and expansion history}.
\newblock Phys. Rev. D \textbf{72}, 043,529 (2005).
\newblock \doi{10.1103/PhysRevD.72.043529}.
\newblock
  {\href{https://arxiv.org/abs/astro-ph/0507263}{{arXiv:astro-ph/0507263}}}

\bibitem{Peebles:1994xt}
P.J.E. Peebles, \emph{Principles of physical cosmology} (1994)

\bibitem{Wang:1998gt}
L.M. Wang, P.J. Steinhardt, {Cluster abundance constraints on quintessence
  models}.
\newblock Astrophys. J. \textbf{508}, 483--490 (1998).
\newblock \doi{10.1086/306436}.
\newblock
  {\href{https://arxiv.org/abs/astro-ph/9804015}{{arXiv:astro-ph/9804015}}}

\bibitem{Huterer:2006mva}
D.~Huterer, E.V. Linder, {Separating Dark Physics from Physical Darkness:
  Minimalist Modified Gravity vs. Dark Energy}.
\newblock Phys. Rev. D \textbf{75}, 023,519 (2007).
\newblock \doi{10.1103/PhysRevD.75.023519}.
\newblock
  {\href{https://arxiv.org/abs/astro-ph/0608681}{{arXiv:astro-ph/0608681}}}

\bibitem{Dvali:2000hr}
G.R. Dvali, G.~Gabadadze, M.~Porrati, {4-D gravity on a brane in 5-D Minkowski
  space}.
\newblock Phys. Lett. B \textbf{485}, 208--214 (2000).
\newblock \doi{10.1016/S0370-2693(00)00669-9}.
\newblock {\href{https://arxiv.org/abs/hep-th/0005016}{{arXiv:hep-th/0005016}}}

\bibitem{Linder:2007hg}
E.V. Linder, R.N. Cahn, {Parameterized Beyond-Einstein Growth}.
\newblock Astropart. Phys. \textbf{28}, 481--488 (2007).
\newblock \doi{10.1016/j.astropartphys.2007.09.003}.
\newblock
  {\href{https://arxiv.org/abs/astro-ph/0701317}{{arXiv:astro-ph/0701317}}}

\bibitem{Zhang:2003ii}
J.~Zhang, L.~Hui, A.~Stebbins, {Isolating geometry in weak lensing
  measurements}.
\newblock Astrophys. J. \textbf{635}, 806--820 (2005).
\newblock \doi{10.1086/497676}.
\newblock
  {\href{https://arxiv.org/abs/astro-ph/0312348}{{arXiv:astro-ph/0312348}}}

\bibitem{Bernstein:2003es}
G.M. Bernstein, B.~Jain, {Dark energy constraints from weak lensing cross -
  correlation cosmography}.
\newblock Astrophys. J. \textbf{600}, 17--25 (2004).
\newblock \doi{10.1086/379768}.
\newblock
  {\href{https://arxiv.org/abs/astro-ph/0309332}{{arXiv:astro-ph/0309332}}}

\bibitem{Ishak:2005zs}
M.~Ishak, A.~Upadhye, D.N. Spergel, {Probing cosmic acceleration beyond the
  equation of state: Distinguishing between dark energy and modified gravity
  models}.
\newblock Phys. Rev. D \textbf{74}, 043,513 (2006).
\newblock \doi{10.1103/PhysRevD.74.043513}.
\newblock
  {\href{https://arxiv.org/abs/astro-ph/0507184}{{arXiv:astro-ph/0507184}}}

\bibitem{Knox:2005rg}
L.~Knox, Y.S. Song, J.A. Tyson, {Distance-redshift and growth-redshift
  relations as two windows on acceleration and gravitation: Dark energy or new
  gravity?}
\newblock Phys. Rev. D \textbf{74}, 023,512 (2006).
\newblock \doi{10.1103/PhysRevD.74.023512}.
\newblock
  {\href{https://arxiv.org/abs/astro-ph/0503644}{{arXiv:astro-ph/0503644}}}

\bibitem{Bertschinger:2006aw}
E.~Bertschinger, {On the Growth of Perturbations as a Test of Dark Energy}.
\newblock Astrophys. J. \textbf{648}, 797--806 (2006).
\newblock \doi{10.1086/506021}.
\newblock
  {\href{https://arxiv.org/abs/astro-ph/0604485}{{arXiv:astro-ph/0604485}}}

\bibitem{Wang:2007fsa}
S.~Wang, L.~Hui, M.~May, Z.~Haiman, {Is Modified Gravity Required by
  Observations? An Empirical Consistency Test of Dark Energy Models}.
\newblock Phys. Rev. D \textbf{76}, 063,503 (2007).
\newblock \doi{10.1103/PhysRevD.76.063503}.
\newblock {\href{https://arxiv.org/abs/0705.0165}{{arXiv:0705.0165}}}
  {[astro-ph]}

\bibitem{Ruiz:2014hma}
E.J. Ruiz, D.~Huterer, {Testing the dark energy consistency with geometry and
  growth}.
\newblock Phys. Rev. D \textbf{91}, 063,009 (2015).
\newblock \doi{10.1103/PhysRevD.91.063009}.
\newblock {\href{https://arxiv.org/abs/1410.5832}{{arXiv:1410.5832}}}
  {[astro-ph.CO]}

\bibitem{Bernal:2015zom}
J.L. Bernal, L.~Verde, A.J. Cuesta, {Parameter splitting in dark energy: is
  dark energy the same in the background and in the cosmic structures?}
\newblock JCAP \textbf{02}, 059 (2016).
\newblock \doi{10.1088/1475-7516/2016/02/059}.
\newblock {\href{https://arxiv.org/abs/1511.03049}{{arXiv:1511.03049}}}
  {[astro-ph.CO]}

\bibitem{DES:2020iqt}
J.~Muir, et~al., {DES Y1 results: Splitting growth and geometry to test
  $\Lambda$CDM}.
\newblock Phys. Rev. D \textbf{103}(2), 023,528 (2021).
\newblock \doi{10.1103/PhysRevD.103.023528}.
\newblock {\href{https://arxiv.org/abs/2010.05924}{{arXiv:2010.05924}}}
  {[astro-ph.CO]}

\bibitem{Ruiz-Zapatero:2021rzl}
J.~Ruiz-Zapatero, et~al., {Geometry versus growth - Internal consistency of the
  flat $\Lambda$CDM model with KiDS-1000}.
\newblock Astron. Astrophys. \textbf{655}, A11 (2021).
\newblock \doi{10.1051/0004-6361/202141350}.
\newblock {\href{https://arxiv.org/abs/2105.09545}{{arXiv:2105.09545}}}
  {[astro-ph.CO]}

\bibitem{Andrade:2021njl}
U.~Andrade, D.~Anbajagane, R.~von Marttens, D.~Huterer, J.~Alcaniz, {A test of
  the standard cosmological model with geometry and growth}.
\newblock JCAP \textbf{11}, 014 (2021).
\newblock \doi{10.1088/1475-7516/2021/11/014}.
\newblock {\href{https://arxiv.org/abs/2107.07538}{{arXiv:2107.07538}}}
  {[astro-ph.CO]}

\bibitem{Mortonson:2008qy}
M.J. Mortonson, W.~Hu, D.~Huterer, {Falsifying Paradigms for Cosmic
  Acceleration}.
\newblock Phys. Rev. D \textbf{79}, 023,004 (2009).
\newblock \doi{10.1103/PhysRevD.79.023004}.
\newblock {\href{https://arxiv.org/abs/0810.1744}{{arXiv:0810.1744}}}
  {[astro-ph]}

\bibitem{Mortonson:2009hk}
M.J. Mortonson, W.~Hu, D.~Huterer, {Testable dark energy predictions from
  current data}.
\newblock Phys. Rev. D \textbf{81}, 063,007 (2010).
\newblock \doi{10.1103/PhysRevD.81.063007}.
\newblock {\href{https://arxiv.org/abs/0912.3816}{{arXiv:0912.3816}}}
  {[astro-ph.CO]}

\bibitem{Vanderveld:2012ec}
R.A. Vanderveld, M.J. Mortonson, W.~Hu, T.~Eifler, {Testing dark energy
  paradigms with weak gravitational lensing}.
\newblock Phys. Rev. D \textbf{85}, 103,518 (2012).
\newblock \doi{10.1103/PhysRevD.85.103518}.
\newblock {\href{https://arxiv.org/abs/1203.3195}{{arXiv:1203.3195}}}
  {[astro-ph.CO]}

\bibitem{Miranda:2017mnw}
V.~Miranda, C.~Dvorkin, {Model-Independent Predictions for Smooth Cosmic
  Acceleration Scenarios}.
\newblock Phys. Rev. D \textbf{98}(4), 043,537 (2018).
\newblock \doi{10.1103/PhysRevD.98.043537}.
\newblock {\href{https://arxiv.org/abs/1712.04289}{{arXiv:1712.04289}}}
  {[astro-ph.CO]}

\bibitem{Raveri:2019mxg}
M.~Raveri, {Reconstructing Gravity on Cosmological Scales}.
\newblock Phys. Rev. D \textbf{101}(8), 083,524 (2020).
\newblock \doi{10.1103/PhysRevD.101.083524}.
\newblock {\href{https://arxiv.org/abs/1902.01366}{{arXiv:1902.01366}}}
  {[astro-ph.CO]}

\bibitem{Hojjati:2013xqa}
A.~Hojjati, L.~Pogosian, A.~Silvestri, G.B. Zhao, {Observable physical modes of
  modified gravity}.
\newblock Phys. Rev. \textbf{D89}(8), 083,505 (2014).
\newblock \doi{10.1103/PhysRevD.89.083505}.
\newblock {\href{https://arxiv.org/abs/1312.5309}{{arXiv:1312.5309}}}
  {[astro-ph.CO]}

\bibitem{Baker:2012zs}
T.~Baker, P.G. Ferreira, C.~Skordis, {The Parameterized Post-Friedmann
  framework for theories of modified gravity: concepts, formalism and
  examples}.
\newblock Phys. Rev. \textbf{D87}(2), 024,015 (2013).
\newblock \doi{10.1103/PhysRevD.87.024015}.
\newblock {\href{https://arxiv.org/abs/1209.2117}{{arXiv:1209.2117}}}
  {[astro-ph.CO]}

\bibitem{Creminelli:2008wc}
P.~Creminelli, G.~D'Amico, J.~Norena, F.~Vernizzi, {The Effective Theory of
  Quintessence: the w<-1 Side Unveiled}.
\newblock JCAP \textbf{0902}, 018 (2009).
\newblock \doi{10.1088/1475-7516/2009/02/018}.
\newblock {\href{https://arxiv.org/abs/0811.0827}{{arXiv:0811.0827}}}
  {[astro-ph]}

\bibitem{Baker:2011jy}
T.~Baker, P.G. Ferreira, C.~Skordis, J.~Zuntz, {Towards a fully consistent
  parameterization of modified gravity}.
\newblock Phys. Rev. \textbf{D84}, 124,018 (2011).
\newblock \doi{10.1103/PhysRevD.84.124018}.
\newblock {\href{https://arxiv.org/abs/1107.0491}{{arXiv:1107.0491}}}
  {[astro-ph.CO]}

\bibitem{Battye:2012eu}
R.A. Battye, J.A. Pearson, {Effective action approach to cosmological
  perturbations in dark energy and modified gravity}.
\newblock JCAP \textbf{1207}, 019 (2012).
\newblock \doi{10.1088/1475-7516/2012/07/019}.
\newblock {\href{https://arxiv.org/abs/1203.0398}{{arXiv:1203.0398}}}
  {[hep-th]}

\bibitem{Gleyzes:2013ooa}
J.~Gleyzes, D.~Langlois, F.~Piazza, F.~Vernizzi, {Essential Building Blocks of
  Dark Energy}.
\newblock JCAP \textbf{1308}, 025 (2013).
\newblock \doi{10.1088/1475-7516/2013/08/025}.
\newblock {\href{https://arxiv.org/abs/1304.4840}{{arXiv:1304.4840}}}
  {[hep-th]}

\bibitem{Gleyzes:2014qga}
J.~Gleyzes, D.~Langlois, F.~Piazza, F.~Vernizzi, {Exploring gravitational
  theories beyond Horndeski}.
\newblock JCAP \textbf{1502}, 018 (2015).
\newblock \doi{10.1088/1475-7516/2015/02/018}.
\newblock {\href{https://arxiv.org/abs/1408.1952}{{arXiv:1408.1952}}}
  {[astro-ph.CO]}

\bibitem{Bloomfield:2012ff}
J.K. Bloomfield, E.E. Flanagan, M.~Park, S.~Watson, {Dark energy or modified
  gravity? An effective field theory approach}.
\newblock JCAP \textbf{1308}, 010 (2013).
\newblock \doi{10.1088/1475-7516/2013/08/010}.
\newblock {\href{https://arxiv.org/abs/1211.7054}{{arXiv:1211.7054}}}
  {[astro-ph.CO]}

\bibitem{Amendola:2013qna}
L.~Amendola, S.~Fogli, A.~Guarnizo, M.~Kunz, A.~Vollmer, {Model-independent
  constraints on the cosmological anisotropic stress}.
\newblock Phys. Rev. \textbf{D89}(6), 063,538 (2014).
\newblock \doi{10.1103/PhysRevD.89.063538}.
\newblock {\href{https://arxiv.org/abs/1311.4765}{{arXiv:1311.4765}}}
  {[astro-ph.CO]}

\bibitem{Noller:2013wca}
J.~Noller, F.~von Braun-Bates, P.G. Ferreira, {Relativistic scalar fields and
  the quasistatic approximation in theories of modified gravity}.
\newblock Phys. Rev. \textbf{D89}(2), 023,521 (2014).
\newblock \doi{10.1103/PhysRevD.89.023521}.
\newblock {\href{https://arxiv.org/abs/1310.3266}{{arXiv:1310.3266}}}
  {[astro-ph.CO]}

\bibitem{Caldwell:2007cw}
R.~Caldwell, A.~Cooray, A.~Melchiorri, {Constraints on a New Post-General
  Relativity Cosmological Parameter}.
\newblock Phys. Rev. \textbf{D76}, 023,507 (2007).
\newblock \doi{10.1103/PhysRevD.76.023507}.
\newblock
  {\href{https://arxiv.org/abs/astro-ph/0703375}{{arXiv:astro-ph/0703375}}}
  {[ASTRO-PH]}

\bibitem{DES:2018ufa}
T.M.C. Abbott, et~al., {Dark Energy Survey Year 1 Results: Constraints on
  Extended Cosmological Models from Galaxy Clustering and Weak Lensing}.
\newblock Phys. Rev. D \textbf{99}(12), 123,505 (2019).
\newblock \doi{10.1103/PhysRevD.99.123505}.
\newblock {\href{https://arxiv.org/abs/1810.02499}{{arXiv:1810.02499}}}
  {[astro-ph.CO]}

\bibitem{Zhang:2007nk}
P.~Zhang, M.~Liguori, R.~Bean, S.~Dodelson, {Probing Gravity at Cosmological
  Scales by Measurements which Test the Relationship between Gravitational
  Lensing and Matter Overdensity}.
\newblock Phys. Rev. Lett. \textbf{99}, 141,302 (2007).
\newblock \doi{10.1103/PhysRevLett.99.141302}.
\newblock {\href{https://arxiv.org/abs/0704.1932}{{arXiv:0704.1932}}}
  {[astro-ph]}

\bibitem{Hu:2007pj}
W.~Hu, I.~Sawicki, {A Parameterized Post-Friedmann Framework for Modified
  Gravity}.
\newblock Phys. Rev. D \textbf{76}, 104,043 (2007).
\newblock \doi{10.1103/PhysRevD.76.104043}.
\newblock {\href{https://arxiv.org/abs/0708.1190}{{arXiv:0708.1190}}}
  {[astro-ph]}

\bibitem{Hojjati:2015ojt}
A.~Hojjati, A.~Plahn, A.~Zucca, L.~Pogosian, P.~Brax, A.C. Davis, G.B. Zhao,
  {Searching for scalar gravitational interactions in current and future
  cosmological data}.
\newblock Phys. Rev. \textbf{D93}(4), 043,531 (2016).
\newblock \doi{10.1103/PhysRevD.93.043531}.
\newblock {\href{https://arxiv.org/abs/1511.05962}{{arXiv:1511.05962}}}
  {[astro-ph.CO]}

\bibitem{Salvatelli:2016mgy}
V.~Salvatelli, F.~Piazza, C.~Marinoni, {Constraints on modified gravity from
  Planck 2015: when the health of your theory makes the difference}.
\newblock JCAP \textbf{1609}(09), 027 (2016).
\newblock \doi{10.1088/1475-7516/2016/09/027}.
\newblock {\href{https://arxiv.org/abs/1602.08283}{{arXiv:1602.08283}}}
  {[astro-ph.CO]}

\bibitem{Mueller:2016kpu}
E.M. Mueller, W.~Percival, E.~Linder, S.~Alam, G.B. Zhao, A.G. Sánchez,
  F.~Beutler, J.~Brinkmann, {The clustering of galaxies in the completed
  SDSS-III Baryon Oscillation Spectroscopic Survey: constraining modified
  gravity}.
\newblock Mon. Not. Roy. Astron. Soc. \textbf{475}(2), 2122--2131 (2018).
\newblock \doi{10.1093/mnras/stx3232}.
\newblock {\href{https://arxiv.org/abs/1612.00812}{{arXiv:1612.00812}}}
  {[astro-ph.CO]}

\bibitem{Daniel:2012kn}
S.F. Daniel, E.V. Linder, {Constraining Cosmic Expansion and Gravity with
  Galaxy Redshift Surveys}.
\newblock JCAP \textbf{1302}, 007 (2013).
\newblock \doi{10.1088/1475-7516/2013/02/007}.
\newblock {\href{https://arxiv.org/abs/1212.0009}{{arXiv:1212.0009}}}
  {[astro-ph.CO]}

\bibitem{Simpson:2012ra}
F.~Simpson, et~al., {CFHTLenS: Testing the Laws of Gravity with Tomographic
  Weak Lensing and Redshift Space Distortions}.
\newblock Mon. Not. Roy. Astron. Soc. \textbf{429}, 2249 (2013).
\newblock \doi{10.1093/mnras/sts493}.
\newblock {\href{https://arxiv.org/abs/1212.3339}{{arXiv:1212.3339}}}
  {[astro-ph.CO]}

\bibitem{Ade:2015rim}
P.A.R. Ade, et~al., {Planck 2015 results. XIV. Dark energy and modified
  gravity}.
\newblock Astron. Astrophys. \textbf{594}, A14 (2016).
\newblock \doi{10.1051/0004-6361/201525814}.
\newblock {\href{https://arxiv.org/abs/1502.01590}{{arXiv:1502.01590}}}
  {[astro-ph.CO]}

\bibitem{Annis:2022xgg}
J.~Annis, J.A. Newman, A.~Slosar, {Snowmass2021 Cosmic Frontier: Report of the
  CF04 Topical Group on Dark Energy and Cosmic Acceleration in the Modern
  Universe}  (2022).
\newblock {\href{https://arxiv.org/abs/2209.08049}{{arXiv:2209.08049}}}
  {[astro-ph.CO]}

\end{thebibliography}


\end{document}